\documentclass[reprint, aps, onecolumn, amsmath, amssymb]{revtex4-2}

\usepackage{graphicx}

\usepackage{newtxtext}
\usepackage{newtxmath}
\usepackage{natbib}
\usepackage{hyperref}
\hypersetup{
    colorlinks = true,
    urlcolor   = blue,
    citecolor  = black,
}

\newcommand{\RomanNumeralCaps}[1]
\linenumbers

\begin{document}

\title{Peeling fingers in an elastic Hele-Shaw  channel}

\author
 {Jo\~ ao V. Fontana}
 \author{Callum Cuttle}
 \author{Draga Pihler-Puzovi{\'{c}}}
 \author{Andrew L. Hazel}
  \author{Anne Juel}
  \email{anne.juel@manchester.ac.uk}

\affiliation
{
Manchester Centre for Nonlinear Dynamics, University of
Manchester, Oxford Road, Manchester M13 9PL, UK
}

\begin{abstract}
	
	Using experiments and a depth-averaged numerical model, we study instabilities of two-phase flows in a Hele-Shaw channel with an elastic upper boundary and a non-uniform cross-section prescribed by initial collapse. Experimentally, we find increasingly complex and unsteady modes of air-finger propagation as the dimensionless bubble speed, $Ca$, and level of collapse are increased, including pointed fingers, indented fingers and the feathered modes first identified by Cuttle \emph{et al.} (\emph{J. Fluid Mech.}, vol. 886, 2020, A20). 
	
	By introducing a measure of the viscous contribution to finger propagation, we identify a $Ca$ threshold beyond which viscous forces are superseded by elastic effects. Quantitative prediction of this transition between `viscous' and `elastic' reopening regimes across levels of collapse establishes the fidelity of the numerical model. In the viscous regime, we recover the non-monotonic dependence on $Ca$ of the finger pressure, which is characteristic of benchtop models of airway reopening. To explore the elastic regime numerically, we extend the depth-averaged model introduced by Fontana \emph{et al.} (\emph{J. Fluid Mech.}, vol. 916, 2021, A27) to include an artificial disjoining pressure which prevents the unphysical self-intersection of the interface.
	
	Using time simulations, we capture for the first time the majority of experimental finger dynamics, including feathered modes. We show that these disordered states continually evolve, with no evidence of convergence to steady or periodic states. We find that the steady bifurcation structure satisfactorily predicts the bubble pressure as a function of $Ca$, but that it does not provide sufficient information to predict the transition to unsteady dynamics which appears strongly nonlinear.

\end{abstract}

\maketitle
%\begin{keywords}
%Fingering instability, Hele-Shaw flows, flow–vessel interactions.
%\end{keywords}

\section{Introduction}
\label{Intro_paper_2}

Thin-film flows confined by elastic boundaries occur in a plethora of applications related to industrial, geophysical and biological processes \citep{FSI}, e.g., from roll coating \citep{Carvalho} to spleenic filtration of red blood cells \citep{viallat}. In microfluidics, there is considerable interest in incorporating thin membranes and other soft boundaries into devices to harness flow-induced functionality and enable passive flow control \citep{Stone, Christov, Hosoi, Vella}. However, many practical microfluidic flows are multiphase, and are thus jointly influenced by elastic boundaries and moving fluid interfaces, which each introduce nonlinearities into otherwise linear flows in the Stokes limit. In this paper, we use a combined approach of experiments and numerical modelling to map out remarkably complex dynamics in a conceptually simple example of such a flow, namely a two-phase displacement flow in a Hele-Shaw channel where the top wall is an elastic sheet. 

Two-phase displacement in a rigid Hele-Shaw channel, a channel whose width is much greater than its depth, is an exemplar of complex interfacial dynamics \citep{Homsy1987, Couder2000, casademunt2004}. When a less viscous fluid, usually air, invades a more viscous fluid, the Saffman--Taylor viscous fingering instability leads to the steady propagation of a single, centred finger \citep{SaffmanTaylor1958}, which can be captured accurately with numerical models \citep{McLeanSaffman1981}, provided the liquid films left behind the advancing finger tip are taken into account \citep{Tabeling1986}. Although this finger is linearly stable \citep{Bensimon1987} for all dimensionless finger speeds, quantified by a capillary number $Ca$, finger instabilities are readily observed under finite-amplitude perturbations imposed either via the geometry \citep{Couder2000} or via the fluid, e.g., by suspending particles \citep{Lindner_2007}. 

In the rigid configuration, injected air displaces resident fluid along the length of the channel. In contrast, the injection of air into a liquid-filled compliant channel inflates its elastic top wall, imposing a reopening channel profile which localises fluid redistribution to a wedge-shaped region ahead of the advancing interface \citep{JuelARFM2018}.

This results in the peeling of the elastic top sheet from the rigid bottom wall at an angle which is set by the fluid-structure interaction \citep{PihlerPeng2015, PengPihler2015, PengLister2019}.
Related peeling scenarios arise in the context of pulmonary airway reopening \citep{Grotberg2001, HeilHazel2011}, where benchtop models have characterised the steady propagation of an air finger into two-dimensional elastic channels under axial tension \citep{Gaver1990, Gaver1996, Jensen2002} and collapsed elastic tubes \citep{HazelHeil2003, JuelHeap2007}. For moderate levels of tube collapse where the top elastic wall does not contact the bottom boundary, reopening takes place either via steady peeling modes, where an increase in pressure drives faster fingers, or steady pushing modes, where the converse is true, \textsl{i.e.}, an increase in pressure drives slower fingers. Transition between these two modes occurs at a critical $Ca_c$ and a yield pressure difference (bubble pressure relative to the pressure in the collapsed channel), which must be exceeded in order for a finger to propagate. The less intuitive pushing mode ($Ca< Ca_c$), which has been found to be unstable in a two-dimensional model \citep{Halpern2005}, follows from the fact that as $Ca$ is reduced, the elastic channel must expand indefinitely to accommodate redistribution of a finite volume of fluid within the ever-thinning films on its walls \citep{HazelHeil2003}. 

It is well established that in a radial Hele-Shaw cell of uniform depth, an elastic top wall leads to the suppression of viscous fingering instabilities \citep{Draga2012PRL,JuelARFM2018}. This is because the interface advances into a convergent channel, where both viscous and surface tension forces act to stabilise interface perturbations \citep{Housseiny2012,  Pihler2018, PengLister2019}. The situation is more complex for finger propagation in our rigid Hele-Shaw channel where the top boundary is an elastic sheet. This is because the presence of side walls and initial channel collapse lead to a cross-section of non-uniform depth, with maximum constriction in the centre of the channel. The reopening mechanics of this system were first explored by \cite{Ducloue2017b}. Remarkably, they found that unsteady finger propagation characterised by complex pattern formation at the finger tip, was supported for sufficiently large $Ca$ across a broad range of initial levels of collapse. For relatively large initial collapse, \cite{Callum2020} identified two apparently persistent modes of finger propagation amongst a host of transient dynamics, which were mediated over a region of bistability by unstable pushing behaviour. These distinct reopening modes were shown to be dominated by viscous and elastic forces, respectively. The elastic reopening mode was associated with small-amplitude fingering perturbations which yielded intricate interfacial patterns referred to as `feathered' modes. However, the nature of these feathered modes remains elusive. 

A modelling framework for fluid-structure interaction flows was proposed by \cite{Fontana2020}, based on a depth-averaged model to describe the propagation of an air finger into a collapsed elasto-rigid channel. They showed that the presence of the elastic wall can lead to interaction between solution branches that are isolated in the rigid channel, thus altering their stability and potentially leading to complex dynamics at higher levels of initial collapse. However, the model was unable to predict the myriad of exotic fingering patterns found experimentally \citep{Ducloue2017b, Callum2020}. In this paper,
we enable the simulation of intricate interfacial patterns by extending the depth-averaged model introduced by \cite{Fontana2020} to include an artificial  disjoining pressure to prevent self-intersection of the interface. We systematically investigate the transition to feathered modes across a range of levels of collapse. Remarkably, we find that they arise after long oscillatory transients which resemble our experimental observations at lower capillary numbers. Furthermore, both experiment and model indicate that the small scale indentations which develop and advect around the finger tip are refined through tip-splitting as the finger propagates. This implies that the feathered modes of propagation are in fact continually evolving, with no evidence that these disordered states converge to steady or periodic states. \cite{Fontana2020} also demonstrated that a thin-film model is fundamental to capturing experimental results both qualitatively and quantitatively. In this paper, we refine their thin-film model to improve the prediction of finger speed as a function of injection flow rate, but we also find that the dynamics are not sensitive to the exact choice of thin-film model.

 The paper is organised as follows. We recall the experimental methodology in \S \ref{Experimental_setup_paper_2} and highlight the essential new features of the numerical model in \S \ref{Model_paper_2}. Results are presented in \S \ref{Results_paper_2}. We introduce a measure of the dimensionless finger speed based on viscous dissipation within the fluid ahead of the finger tip in \S \ref{darcys_factor_paper_2}, which enables us to identify a threshold value of $Ca$ beyond which viscous forces are superseded by elastic effects. Quantitative prediction of this transition between `viscous' and `elastic' reopening regimes across levels of collapse establishes the fidelity of the numerical model. We show in \S \ref{slightly_collapsed_paper_2} that in the viscous regime, we recover the non-monotonic dependence on $Ca$ of the finger pressure, which is characteristic of benchtop models of airway reopening. In \S \ref{multiple_modes_paper_2}, we explore the elastic regime and the transition to feathered states as a function of $Ca$. We capture feathered states numerically for the first time in \S \ref{t_evo_unstable_modes_paper_2}. A detailed analysis of the steady bifurcation structure in \S \ref{p_Ca_relations_paper_2} reveals that steady numerical solutions match the bounding envelopes of unsteady fingers in the elastic regime and, thus, satisfactorily predict the bubble pressure. However, the steady bifurcation diagram does not inform the transition to unsteady dynamics, which appears strongly nonlinear.

\vspace{1cm}

\section{Experimental set-up}
\label{Experimental_setup_paper_2}

\begin{figure}
	\includegraphics[scale=0.42]{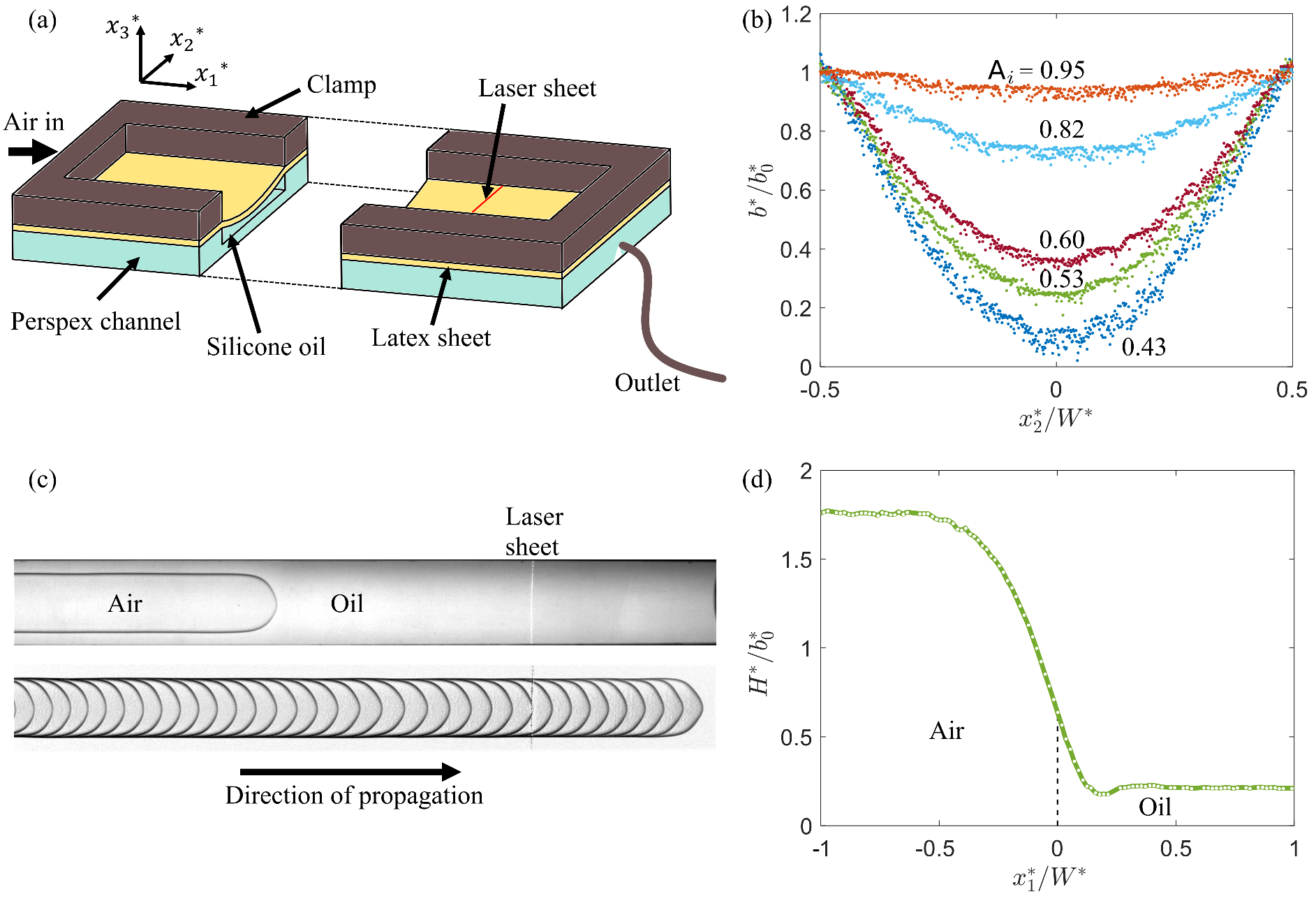}
	\caption{(a) Schematic diagram of the experimental set-up. Adapted from~\cite{Callum2020}. (b) Initial elastic-sheet profiles across the width $W^*$ of the channel. Membrane height $b^*$, measured by the laser sheet depicted in (a), as a function of the scaled lateral coordinate $x_2^*/W^*$ is normalised by the channel depth $b_0^*$. Labels refer to $A_i$, a measure of initial collapse. (c) Top: image of air finger within the experimental region of interest (ROI). This image was captured by the top-view camera during the reopening of a channel with $A_i=0.53$ at flow rate $Q^*=90$~ml/min. Bottom: composite image produced by overlaying successive images from the experiment, taken at time intervals of $0.17$~s. Time increases from left to right. (d) Instantaneous elastic-sheet height $H^*$ mid-way between the channel walls ($x_2^*=0$) as a function of axial coordinate $x_1^*$ during reopening [taken from the same experiment as (c)]. The dashed line indicates the position of the air-oil interface at the tip of the air finger.}
	\label{Exp_set_up_fig}
\end{figure}

We performed experiments in a rigid Hele-Shaw channel topped with an elastic sheet, shown schematically in figure~\ref{Exp_set_up_fig}(a). This experimental set-up was previously described by \cite{Callum2020}. A channel of length $60$~cm, width $W^{*}=30\pm0.02$~mm and depth $b_0^{*}=1.05\pm0.01$~mm was precision-milled into a block of Perspex, achieving a 10 $\mu$m roughness along the base. For the top boundary, we used a Latex sheet (Supatex) with Young's modulus $E^{*}=1.44\pm0.05$~MPa, Poisson's ratio $\nu=0.5$ and thickness $h^{*}=0.46\pm0.01$~mm. A uniform pre-stress was imposed on the elastic sheet, directed across the width of the channel, by hanging evenly distributed weights (totalling $3.03$~kg) along one long edge of the sheet. This pre-stress was maintained by clamping the elastic sheet in place using an aluminium frame secured with 11 evenly-spaced G-clamps along each long edge and a single bolt along each short edge. The constitutive relation between transmural pressure difference across the elastic sheet and level of collapse, which was measured in the experimental channel under static conditions, matches numerical simulations performed using a pre-stress that is $30\%$ of the value predicted from the sheet dimensions and applied weight (see appendix \ref{appendix_a}). This loss of applied pre-stress is expected, due to unavoidable slippage of the elastic sheet during the clamping procedure. However, the reproducibility and consistency of the experimental results suggest that the pre-stress was uniform and remained constant once the elastic sheet was clamped in place.

To set the initial conditions of each experiment, the channel was filled with silicone oil of dynamic viscosity $\mu^{*}=0.099$~Pa.s, surface tension $\gamma^{*}=21$~mN/m and density $\rho^{*}=973$~kg/m${^3}$ at laboratory temperature $T^{*}=21\pm1$~$^{\circ}$C. The inlet was then closed and oil was allowed to drain from the outlet of the channel, via a length of flexible tubing open to the atmosphere at one end. By adjusting the height of the outlet of the tubing relative to the channel we could set the hydrostatic pressure difference between the channel and the atmosphere. At equilibrium, this hydrostatic pressure difference is equal to $p^*_\mathrm{trans}$, the transmural pressure difference across the elastic sheet, which acts to deform the sheet. Hence, we were able to control the extent to which the elastic sheet was initially collapsed. We measured the initial collapse using a laser sheet projected to a line across the width of the channel and imaged this line at an oblique angle using a camera of resolution $22.9\pm0.02$~pixels/mm (the same camera was used for visualising sheet deformation during the reopening experiment, see below). The resulting profiles of the sheet in terms of the local depth $b^{*}$ of the channel as a function of the lateral coordinate $x_2^{*}$ are shown in figure~\ref{Exp_set_up_fig}(b). We quantify the extent of initial collapse with the parameter $A_i$, which is defined as the ratio of cross-sectional area under the sheet to the uncollapsed area $W^{*}b_0^{*}$. Hence, for an uncollapsed channel $A_i=1$. By $A_i=0.36$, the elastic sheet sags sufficiently to contact the centreline of the bottom wall, which we refer to as opposite wall contact. We studied channels collapsed within the range $0.43\pm0.02  \le A_i \le 0.95\pm0.02$, as detailed in figure~\ref{Exp_set_up_fig}(b). 

The channel was reopened by injecting air at a constant volumetric flow rate $Q^{*}$, which varied within the range $5 \le Q^{*} \le 330$~ml/min, resulting in a long continuous finger of air. The finger propagated along the length of the channel, displacing oil and parting the channel walls. Flow was imposed using a syringe pump (KD Scientific) fitted with Gastight syringes (Hamilton), with a small precursor bubble ($<1$~mL) injected immediately before each experiment to set the initial conditions. The pressure $p_\mathrm{b}^{*}$ of the air finger was measured with a differential pressure sensor (Honeywell, $\pm$~5"~H2O) with $\pm$1~Pa resolution. Over a $250$~mm length of the channel, which was chosen as the region of interest (ROI) for our measurements, we recorded constant air-bubble pressure traces to within a typical tolerance of $10$~Pa. 

The air finger was imaged from above by a second camera ($4.8\pm0.1$~pixels~mm$^{-1}$) recording images at fixed rates of $0.33$--$30$ frames per second (f.p.s.) depending on $Q^{*}$ and $A_i$. The channel was back-lit by LED lights diffused through opalescent acrylic. An example of a top-view image is shown in the upper panel of figure~\ref{Exp_set_up_fig}(c). The tip of the air finger was located in each frame using image analysis routines (MATLAB 2016a), which allowed us to calculate the instantaneous axial speed $u_\mathrm{f}^*$ of the finger. To illustrate the time-evolution of the finger over an experiment, we will refer to composite images of the kind shown in the lower panel of figure~\ref{Exp_set_up_fig}(c), which are generated by overlaying background-subtracted images from successive frames of an experiment. In figure~\ref{Exp_set_up_fig}(c) the finger shape is steady over the entire ROI and, since the interfaces are equally spaced at fixed time intervals, the finger advances at a constant speed.

The deflection of the elastic sheet during reopening was measured by recording images of the laser sheet (at $40$--$160$~f.p.s.), which was located near the end of the ROI, see figure~\ref{Exp_set_up_fig}(c). Measurements of the height $H^*$ of the elastic sheet mid-way between the walls of the channel along with knowledge of the finger speed $u_\mathrm{f}^*$ allowed us to reconstruct the profile of the elastic sheet around the finger tip, as shown, e.g., in figure~\ref{Exp_set_up_fig}(d). Ahead of the finger tip (located at axial coordinate $x_1^*=0$) the channel is still collapsed, while behind it the channel is inflated. Hence, the air finger advances into a tapered region behind a wedge-shaped volume of oil.

\section{Model}
\label{Model_paper_2}

We extend the modelling framework developed by \cite{Fontana2020} which couples depth-averaged lubrication equations for the fluid flow to the F{\"o}ppl--von K\'{a}rm\'{a}n plate equations with in-plane pre-stress for the elastic sheet. The original model was developed in a frame of reference moving with the axial velocity of the finger tip so that steady solutions represented fingers propagating steadily in the frame of the laboratory. This model was implemented numerically in {\tt oomph-lib} \citep{HeilHazel2006} to calculate steady solutions, evaluate their linear stability and perform time simulations. Compared to the model of \cite{Fontana2020}, the new model includes an additional artificial disjoining pressure, which has been implemented to prevent the unphysical self-intersection of air-finger interfaces when deep indentations develop during time-simulations, see \S\ref{disjoin_paper_2}. Furthermore, the effect of different approximations for the liquid films separating the finger from the walls of the channel in comparison with experiments has been analyzed.

\subsection{Governing equations}
\label{governing_sub_sec_paper_2}

We define a frame of reference in Cartesian coordinates $(x_{1}^{*},x_{2}^{*},x_{3}^{*})$ moving with the instantaneous axial speed of the finger $u_\mathrm{f}^{*}(t^{*})$. The coordinate $x_{1}^{*}$ spans the channel length, $x_{2}^{*}$ spans the channel width, while $x_{3}^{*}$ is the out-of-plane (height) coordinate. The domain of the channel is bounded so that: $-L_{\mathrm{up}}^*<x_{1}^{*}<L_{\mathrm{down}}^*$, $-W^{*}/2 \leq x_{2}^{*} \leq W^{*}/2$ and $0 \leq x_{3}^{*} \leq b^{*}(x_{1}^{*},x_{2}^{*},t^{*})$, where $ b^{*}(x_{1}^{*},x_{2}^{*},t^{*})$ is the local height of the channel and $L^*=L^*_{\mathrm{up}}+ L^*_{\mathrm{down}} = 25 W^*$ is the channel length. Throughout this manuscript, dimensional quantities are always starred, while dimensionless ones appear unstarred.

We use the channel width, $W^{*}$, to non-dimensionalise the in-plane coordinates $(x_{1},x_{2})$, and the undeformed channel height, $b^{*}_{0}$, for the out-of-plane coordinate $x_{3}$. These define the aspect ratio $\alpha = W^{*}/ b^{*}_{0}$ of the channel. The displacement in the elastic sheet is non-dimensionalised using the in-plane length scale and the fluid pressure is non-dimensionalised using $\mathcal{P}^{*}=12\mu^{*} \alpha^{2} / \mathcal{T}^{*}$, where $ \mathcal{T}^{*}$ is the time scale defined as $\mathcal{T}^{*}=W^{*2}b_{0}^{*}/Q^{*}$. Finally, in-plane velocities are non-dimensionalised by $\mathcal{U}^{*} = W^{*}/\mathcal{T}^{*}$.

Assuming incompressibility, we apply the Reynolds lubrication approximation to the Stokes equation, defined in the frame moving with instantaneous velocity $u_\mathrm{f}(t)=u_\mathrm{f}^{*}(t) \mathcal{T}^{*}/W^{*}$, which yields the depth-averaged governing equation for the fluid pressure $p$:
\begin{equation}
\frac{\partial b}{\partial t} - u_\mathrm{f} \frac{\partial b}{\partial x_{1}}  = \frac{\partial}{\partial x_{\beta}} \left( b^{3} \frac{\partial p}{\partial x_{\beta}} \right),
\label{governing_eq}
\end{equation}
where we use the summation convention and the index $\beta$, and all subsequent Greek indices, takes the values 1 and 2. The unknown speed $u_\mathrm{f}(t)$ is determined by enforcing that the maximum $x_{1}$ coordinate of the interface (i.e., the $x_{1}$ coordinate of the finger tip $x_{\mathrm{1,tip}}$) is set to zero.

We use the F{\"o}ppl--von K\'{a}rm\'{a}n equations \citep{landaulifshitz}
\begin{equation}
\left(\frac{\partial^{2}}{\partial x_{\beta}\partial
	x_{\beta}}\right)
\left(\frac{\partial^{2}}{\partial x_{\gamma}\partial
	x_{\gamma}}\right)w - \eta \frac{\partial}{\partial x_{\gamma}} \left( \sigma_{\beta\gamma} \frac{\partial w}{\partial x_{\beta}} \right) = P, \quad
\frac{\partial \sigma_{\beta\gamma}}{\partial x_{\gamma}} = 0,
\label{Fvk}
\end{equation}
as the governing equations to determine the in-plane and out-of-plane displacements of the elastic sheet, $(v_{1},v_{2})$ and $w$, respectively. The pressure load $P^{*}$ on the elastic sheet is non-dimensionalised using the bending modulus $K^{*} = \frac{E^{*} h^{*3}}{12(1-\nu^2)}$ so that $P=P^{*}W^{*3} / K^{*}$. The parameter $\eta = 12(1-\nu^2) \left( \frac{W^{*}}{h^*} \right) ^2$ describes the relative importance of the in-plane and bending stresses. Finally, the in-plane components of the stress tensor, $\sigma_{\beta \gamma}$ are
\begin{equation}
\sigma_{11} = \frac{\left( \epsilon_{11} +\nu \epsilon_{22} \right)}{1-\nu^2},\quad
\sigma_{22} = \sigma_{22}^{(0)} + \frac{\left( \epsilon_{22} +\nu \epsilon_{11} \right)}{1-\nu^2}, \quad
\sigma_{12} = \sigma_{21} =  \frac{\epsilon_{12}}{1+\nu},
\label{Stress}
\end{equation}
where the in-plane strain is given by
\begin{equation}
\epsilon_{\beta\gamma} =\frac{1}{2} \left( \frac{\partial v_{\beta}}{\partial x_{\gamma}} + \frac{\partial v_{\gamma}}{\partial x_{\beta}} \right) + \frac{1}{2}\frac{\partial w}{\partial x_{\beta}}\frac{\partial w}{\partial x_{\gamma}},
\label{strain}
\end{equation}
and a non-zero pre-stress component $\sigma_{22}^{(0)}$ is introduced to mimic the clamping procedure performed in the experiment which tensions the sheet in the $x^*_2$ direction. The pre-stress components $\sigma_{11}^{(0)}$ and $\sigma_{12}^{(0)}$ are fixed as zero. However, as mentioned in \S \ref{Experimental_setup_paper_2} the experimental value of $\sigma_{22}^{(0)}$ is not known accurately, so, following \cite{Fontana2020}, we estimate it by matching the constitutive relation of the experimental channel to numerical simulations. Details of this procedure are presented in appendix \ref{appendix_a}.

The equations governing the fluid pressure (\ref{governing_eq}) and elastic sheet deformation (\ref{Fvk}) are coupled in two distinct ways. First, the height $b$ of the channel is determined by the out-of-plane displacement $w$ of the elastic sheet:
\begin{equation}
b(x_{1},x_{2},t) = 1 + \alpha w.
\label{eq:channel_h}
\end{equation}
Secondly, the pressure load $P$ on the elastic sheet depends on the pressure $p$ of the fluid and the pressure $p_\mathrm{b}$ of the air finger:
\begin{equation}
P = \mathcal{I}p_\mathrm{b} \ \ \ \ \mbox{in} \ \ \Omega_\mathrm{air},\quad\quad\quad
P =\mathcal{I}p \ \ \ \ \mbox{in} \ \ \Omega_\mathrm{fluid},
\label{p_FSI}
\end{equation}
where the non-dimensional fluid-structure interaction parameter
\begin{equation}
\mathcal{I}=\frac{144\mu^* \alpha^3 (1-\nu^2) Q^*}{E^* h^{*3}}
\label{FSI}
\end{equation}
provides a measure of the typical viscous stresses in the fluid relative to the bending stress of the elastic sheet. As $\mathcal{I} \to 0$ the elastic sheet becomes rigid and the governing equations (\ref{governing_eq}) and (\ref{Fvk}) decouple.

\subsection{Boundary conditions}
\label{sussec:BC}

At the side walls of the channel, the boundary conditions are non-penetration of fluid and a clamped elastic sheet: 
\begin{equation}
\frac{\partial p}{\partial x_{2}} = 0, \ \ \ \ v_{1}=0, \ \ \ \ v_{2}=0, \ \ \ \ \ w=0, \ \ \ \ \frac{\partial w}{\partial x_{2}}=0, \quad\mbox{at}\quad x_{2}= \pm 0.5.
\label{side_wall_bc}
\end{equation}

\begin{figure}
    \includegraphics[scale=0.26]{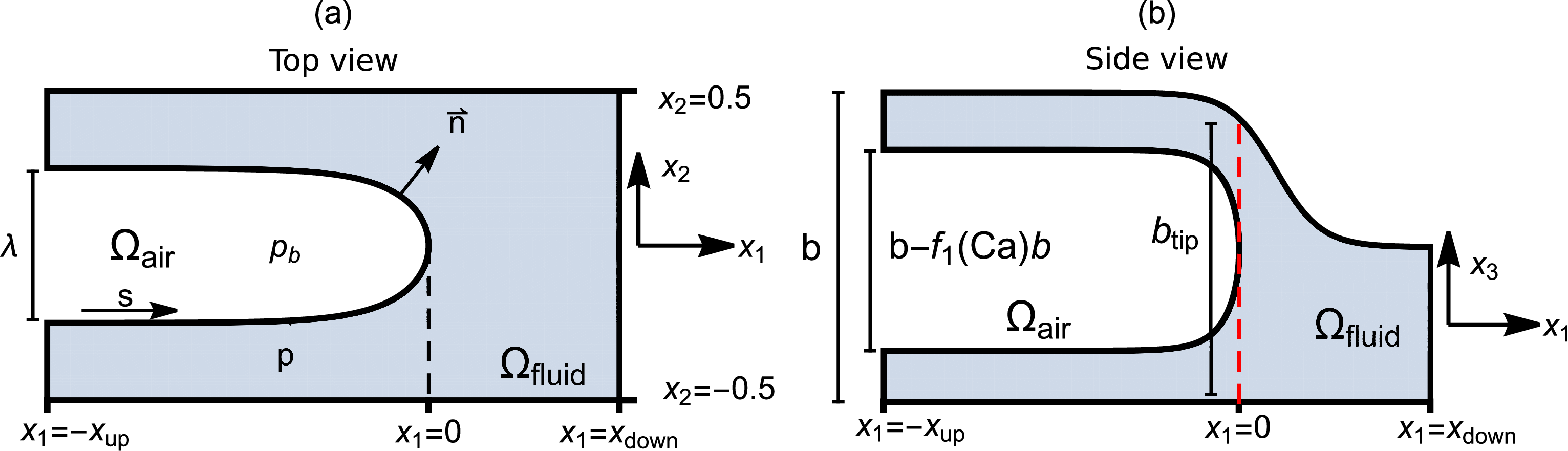}
	\caption{(a) Numerical domain in a frame of reference moving at the velocity of the finger tip. The rigid side walls are located at $x_{2}=0.5$ and $x_{2}=-0.5$. The domain is truncated at $x_{1}=-x_{\mathrm{up}}$ upstream and at $x_{1}=x_{\mathrm{down}}$ downstream. The $x_{1}$ coordinate of the finger tip is fixed at $0$ and the finger width is $\lambda$. (b) Sketch of the transverse view of the channel along the line $x_{2}=x_{\mathrm{2,tip}}$ that crosses the finger tip. The thickness of the fluid film is $f_{1}(Ca)b$ and the thickness of the air finger is $b-f_{1}(Ca)b$. The height of the elastic sheet at the finger tip is denoted by $b_{\mathrm{tip}}=b(x_{\mathrm{1,tip}},x_{\mathrm{2,tip}})$.}
	\label{Numerical_domain}
\end{figure}

The numerical domain is truncated in the axial direction at the upstream ($x_{1} = -x_\mathrm{up}$) and downstream ($x_{1}=x_\mathrm{down}$) ends, see figure \ref{Numerical_domain}(a). We set $x_\mathrm{up}=10$ and $x_\mathrm{down}=15$, and checked that  the computed solutions are not affected by increasing the length of the domain. Following \cite{HazelHeil2003}, at these truncated boundaries, we impose the conditions	
\begin{eqnarray}
v_{1}=0, \ \ \ v_{2}=0, \ \ \ \frac{\partial w}{\partial x_{1}}=0, \ \ \ \ \frac{\partial p}{\partial x_{1}}=0 \ \ \ \ &\mbox{at}& \ \ x_{1}=-x_{\mathrm{up}}, \nonumber \\
v_{1}=0, \ \ \ v_{2}=0,
\ \ \ \frac{\partial w}{\partial x_{1}}=0, \ \ \ \frac{\partial
	p}{\partial x_{1}}= G   \ \ &\mbox{at}& \ \ x_{1}=x_{\mathrm{down}}.
\label{truncated_BC}
\end{eqnarray}
These mean that far away from the finger tip, all disturbances should decay. We allow the upstream pressure gradient $G$ to be an unknown, which we determine by imposing that the fluid flux at the downstream boundary is equal to the flux at $x_{1} \to \infty$:
\begin{equation}
\int_{-0.5}^{0.5}  \left( -b^3G -bu_\mathrm{f} \right)|_{x_{1} = x_{\mathrm{down}}}  dx_{2} = -A_{i}u_\mathrm{f},
\label{Flux_ahead}
\end{equation}
where $A_{i}$ is the initial level of collapse defined in \S \ref{Experimental_setup_paper_2}.

\subsubsection{Effect of fluid films left behind the advancing interface}
\label{fluid_film_effect_paper_2}

Following \cite{PengPihler2015} and \cite{Fontana2020}, we incorporate into the boundary conditions at the air-finger interface the effects of fluid films left behind the advancing interface. The kinematic boundary condition is given by
\begin{equation}
\left( b-f_{1}(Ca)b \right)\left[\frac{\partial \textbf{R}}{\partial t}
+ u_\mathrm{f} \mbox{\boldmath$e$}_{1}\right]\cdot \textbf{n} =
-b^{3} \frac{\partial p}{\partial x_{\alpha}} n_{\alpha}\ \ \mbox{on} \ \ \partial \Omega_\mathrm{air},
\label{K_bc}
\end{equation}
where $\textbf{R}(s,t)$ is the position of the advancing air-fluid interface in the moving frame, parameterized by the coordinate $s$, and $\textbf{n}$ is the in-plane outer unit normal vector to the interface, see figure \ref{Numerical_domain}. The dynamic boundary condition is given by
\begin{equation}
\Delta p=p|_{\partial \Omega_\mathrm{air}} - p_\mathrm{b}=-\frac{u_\mathrm{f}}{12\alpha^2 Ca} \left( \kappa + \alpha\frac{2}{b}f_{2}(Ca) \right),
\label{D_bc}
\end{equation}
where $\kappa$ is the in-plane curvature of the air-finger interface and $Ca=\mu^{*}u_\mathrm{f}^{*} / \gamma^{*}$ is the capillary number. Note that the capillary number is based on the instantaneous velocity of the finger tip. This means that, for unsteadily propagating fingers $Ca$ is time-dependent. The functions $f_{1}(Ca)$ and $f_{2}(Ca)$ incorporate the effect of the fluid films into the model. We use the expressions
\begin{equation}
f_{1}(Ca)=\frac{Ca^{2/3}}{0.76+2.16\,Ca^{2/3}}
\label{f_1}
\end{equation}
and 
\begin{equation}
f_{2}(Ca)=1+\frac{Ca^{2/3}}{0.26+1.48\,Ca^{2/3}} + 1.59\,Ca,
\label{f_2}
\end{equation}
which \cite{PengPihler2015} introduced and validated for use in elastic cells. The effects of fluid films are removed if we set $f_{1}(Ca)=0$ and $f_{2}(Ca)=1$. \cite{PengPihler2015} assumed that the thickness of the fluid film is set at the finger tip position $\Pi_\mathrm{tip} =  (x_{\mathrm{1,tip}} , x_{\mathrm{2,tip}})$, where it is formed. This means that, in their model, the fluid film has a uniform thickness $f_{1}(Ca)b_\mathrm{tip}$ at every point $(x_{1},x_{2})$ of the air finger, where $b_\mathrm{tip} = b(x_{\mathrm{1,tip}} , x_{\mathrm{2,tip}})$. This choice differs from the assumption made by \cite{Fontana2020}, where the thickness of the fluid film was a constant proportion of the channel height $f_{1}(Ca)b(x_{1},x_{2})$ at each point $(x_{1},x_{2})$ of the air finger. We refer to these assumptions as uniform and non-uniform film thickness models. In this paper, we follow \cite{Fontana2020} and assume a non-uniform film thickness model. In appendix \ref{appendix_b} we compare both assumptions with experiments and, in addition, show that the relationship between bubble pressure and capillary number predicted by the model remains the same under both assumptions. The imposed flow rate necessary to reach a given capillary number, however, depends on the assumption made for the film thickness. In fact, both assumptions are simplifications of the three-dimensional fluid rearrangement that takes place within the films; and three-dimensional Stokes simulations would be needed to accurately capture the distribution of fluid films required for detailed quantitative agreement with experiment. Moreover, we find that the behaviour of the fluid in these films does not affect the overall qualitative dynamics.

\subsubsection{Disjoining pressure}
\label{disjoin_paper_2}

\begin{figure}
	\centering
	\includegraphics[scale=0.26]{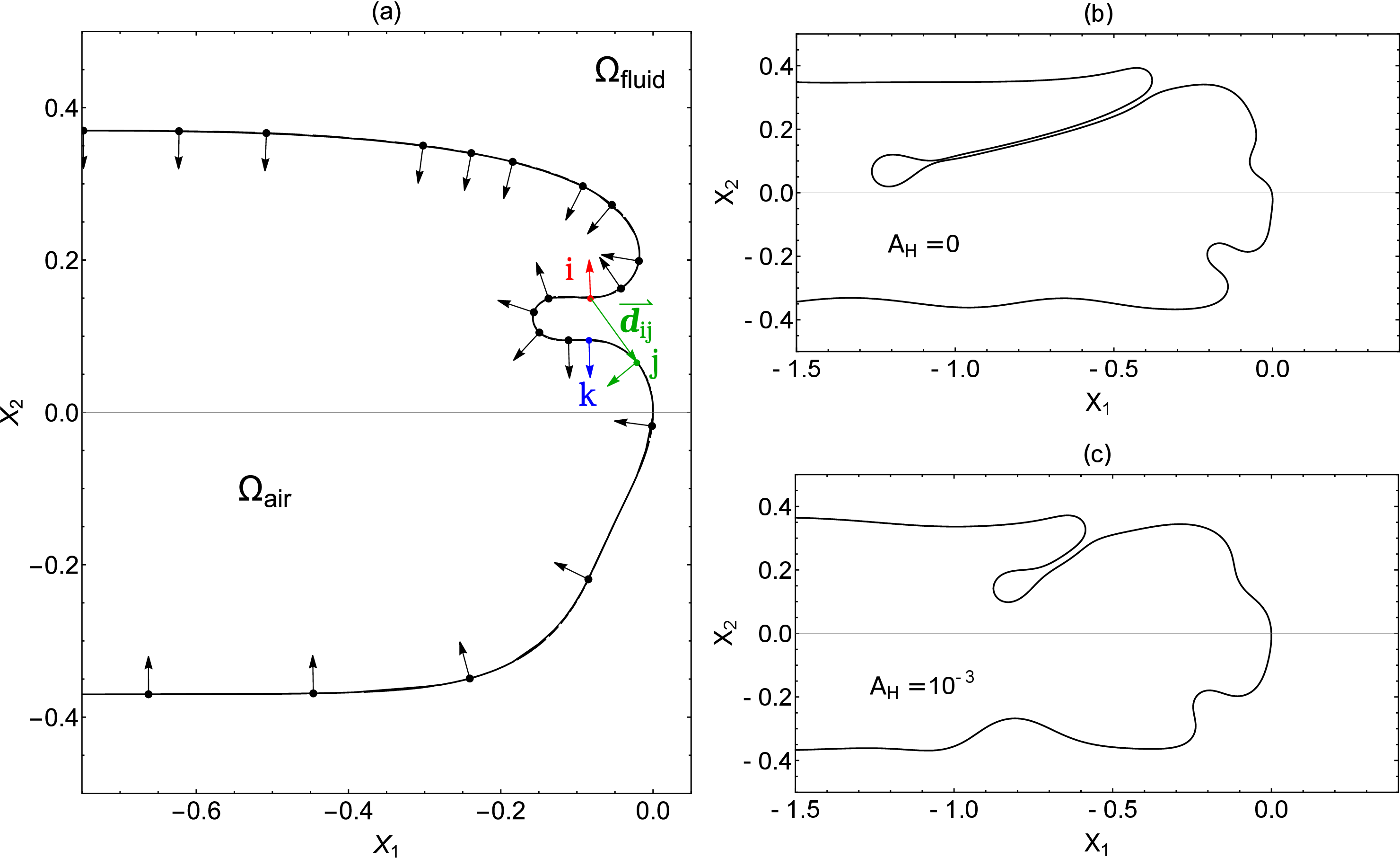}
	\caption{(a) Example of a numerical time-dependent solution where the air-finger interface develops an indentation. One in every 25 nodes is shown with circles on the interface along with their normal vector. The magnitude of the vector $\vec{d}_{ij}$ corresponds to the distance between the points $i$ and $j$. At the point $i$, the interface-interface distance $d$ is equal to $|\vec{d}_{ik}|$. (b) Example of an unsteady simulation where the finger develops an indentation in the absence of disjoining pressure. (c) The same unsteady simulation as in (b), and at the same time, but with the addition of disjoining pressure.}
	\label{disjoin_figure}
\end{figure}

\cite{Fontana2020} found that time simulations at high $Ca$ and low $A_{i}$ produced unsteady fingers with small scale indentations originating near the finger tip. These indentations could grow into clefts sufficiently deep and narrow that they led to the self-intersection of the air-finger interface, thus terminating the simulation. However, self-intersection of the air-finger interface was never observed in the experiments presented by \cite{Ducloue2017b}, \cite{Callum2020} and in this paper. 
It is most likely that the self-intersection in the model is a result of the approximations made when simplifying the three-dimensional liquid-film dynamics. In order to allow time simulations to continue beyond the point of self-intersection, rather than moving to three-dimensional calculations, we prevent self-intersection by adding an artificial repulsive disjoining pressure $p_\mathrm{d}$ to our model.
We choose the form 
\begin{equation}
p_\mathrm{d} = A_{H} \left( \frac{\lambda_{d}}{d} \right)^{3},
\label{disjoining_p}
\end{equation}
based on the disjoining pressure in thin films introduced by
\cite{derjaguin1955}. The constant $A_{H}$ gives the strength of the
repulsive interaction and $\lambda_{d}$ is the length scale of the
interaction. The interface-interface distance $d$, see depiction in
figure \ref{disjoin_figure}(a), was calculated as the distance between
a pair of points on the interface, one on each side of the
indentation. Each interface point was paired with the point with the
shortest distance from it and with the additional constraint that the
unit normal vectors to the interface at these two points have a
negative inner product. This constraint ensures that the points are
always localized at opposite sides of the indentation. Once a pair is
assigned for each point at the interface, we calculate the local value
of the disjoining pressure $p_\mathrm{d}$ for that point. In order to
identify the pairs of points located at indentations, we map all pairs
in the entire finger interface. It is possible to show that for the
pairs of points that are not on an indentation (e.g., a pair of points
located far behind the finger tip, on opposite sides of the finger
width), the distance $d$ is always of the order of the finger width
(between $0.4$ and $0.7$), while for pairs that are on an indentation,
the distance $d$ is $0.05$ or smaller. For
  sufficiently large values of $d$, $p_\mathrm{d}$ becomes negligible
  and, in order to speed up the calculations by reducing the length of the interface that must be
  scanned,  we set the value of
  $p_\mathrm{d}$ to zero for points separated by a distance greater than a
  cutoff value of $0.1$. For our chosen values of $A_{H}$ and
  $\lambda_{d}$, see below, $p_{d} < 10^{-6}$ when $d > 0.1$. The overall effect of this implementation of disjoining pressure is a short ranged repulsive interaction that only acts to prevent self-intersection on interface indentations.

The disjoining pressure was added to the dynamic boundary condition (\ref{D_bc}), resulting in a new condition
\begin{equation}
\Delta p=p|_{\partial \Omega_\mathrm{air}} - p_\mathrm{b} - p_\mathrm{d}.
\label{D_bc_disjoining}
\end{equation}
We used trial and error to select the parameter values $A_{H}=10^{-3}$ and $\lambda_{d}=10^{-2}$, so that the disjoining pressure prevents self-intersection of the interface during time-dependent calculations but only affects the dynamics when the interface is on the verge of self-intersection. For steady state simulations, we set $p_\mathrm{d}=0$.

Figure \ref{disjoin_figure} shows a comparison between time-simulations without (figure \ref{disjoin_figure}(b)) and with (figure \ref{disjoin_figure}(c)) the addition of a disjoining pressure. Both simulations are shown for the same time step and simulated with the same parameters and initial conditions. In the time-step following the snapshot of figure \ref{disjoin_figure}(b), the interface self-intersects, thus terminating the numerical simulation, whereas this does not occur in figure \ref{disjoin_figure}(c). There, the addition of $p_\mathrm{d}$ reduces the depth of the indentation and increases the distance $d$ between the sides of the indentation. This brings unsteady finger dynamics in the simulations closer to the experiments, where indentations reduce in depth as they are advected to the side of the finger before eventually disappearing; see \S \ref{t_evo_unstable_modes_paper_2}.
However, 3D Stokes simulations would be necessary to fully capture the interface-interface interaction in the region of indentations. In the case of deep indentations, the separation $d^{*}$ can be as small as the unperturbed channel gap $b^{*}_{0}$, making the lubrication approximation invalid.

\subsection{Numerical implementation}
\label{numerical_implementation_paper_2}

The numerical implementation of our model followed that of \cite{Fontana2020}, with the addition of the disjoining pressure to the fluid dynamic boundary condition and the possibility to choose between fluid film correction of uniform and non-uniform thickness.

In order to calculate steadily propagating modes (travelling-wave solutions), we set all time-derivatives in the governing equations and boundary conditions to zero. Additionally, we imposed fixed values of the initial level of collapse $A_{i}$ and another global variable which could be either the bubble pressure $p_\mathrm{b}$, capillary number $Ca$ or flow rate $Q^{*}$. Different choices were made for the second controlled variable according to the parameter continuation between solutions branches in the bifurcation diagram to be calculated. Once we had steady solutions, we analysed the linear stability of these solutions at fixed $Q^{*}$ and $A_{i}$ for consistency with the experiment. The solution of the discrete generalised eigenproblem was obtained via the Anasazi solver from Trilinos \citep{Heroux2005}. We computed the 60 most unstable eigenvalues.

Lastly, in order to assess the non-linear stability of the steady solutions, we conducted time simulations for fixed values of the initial level of collapse and flow rate. We used a steady solution as initial condition to which we applied a time-decaying perturbation to the pressure jump across the interface:

\begin{equation}
\Delta p=p|_{\partial \Omega_\mathrm{air}} - p_\mathrm{b} - p_\mathrm{d} - \delta p,
\label{D_bc_disjoining_perturb}
\end{equation}
where the perturbation takes the form
\begin{equation}
\delta p = -\delta p_0  \left( \frac{2t}{t_p} \right)e^{-(t/t_p)}e^{-((x_2 -x_0)/\lambda_p)^2}.
\label{pressure_perturbation}
\end{equation}
This perturbation creates a dimple on a length scale $\lambda_{p}$ centred on an interfacial point with coordinate $x_{2}=x_{0}$. This artificial extra pressure is zero at $t=0$, reaches the maximum $\delta p_{0}$ and decays to zero for $t \gg t_{p}$. We have chosen this particular form of perturbation in order to be able to apply it consistently for all values of parameters used. We set a non-zero value for $x_{0}$, so we prescribe a perturbation  asymmetric about the centreline of the channel, thus avoiding restricting time-simulations to symmetric fingers. In our simulations, we used the following parameters: $\lambda_{p}=0.035$, $x_{0}=0.05$, $t_{p}=0.04$ and $\delta p_{0}=0.2p_\mathrm{b}$.

\section{Results}
\label{Results_paper_2}

We report experiments and numerical simulations of bubble propagation in a  collapsed channel, see figure \ref{Exp_set_up_fig}, for different values of $A_i$ and the imposed air flow rate $Q^*$. Opposite wall contact is avoided by keeping the level of collapse $A_i$ above the value $0.36$. We present results in terms of the capillary number, $0<Ca<1.3$, which provides a measure of the dimensionless finger velocity and is not sensitive to the choice of film model as discussed in \S \ref{fluid_film_effect_paper_2}. The choice of  experimental parameters listed in \S \ref{Experimental_setup_paper_2} means that the fluid-structure interaction parameter $\mathcal{I}$ varies only with $Q^{*}$ in the range $0<\mathcal{I}<1.5 \times 10^4$, and that the other non-dimensional parameters are fixed at $\alpha = 28.6$ and $\eta \sim 40,000$.

\subsection{Comparison with the flow into a rigid wedge}
\label{darcys_factor_paper_2}

\begin{figure}
	\includegraphics[scale=0.60]{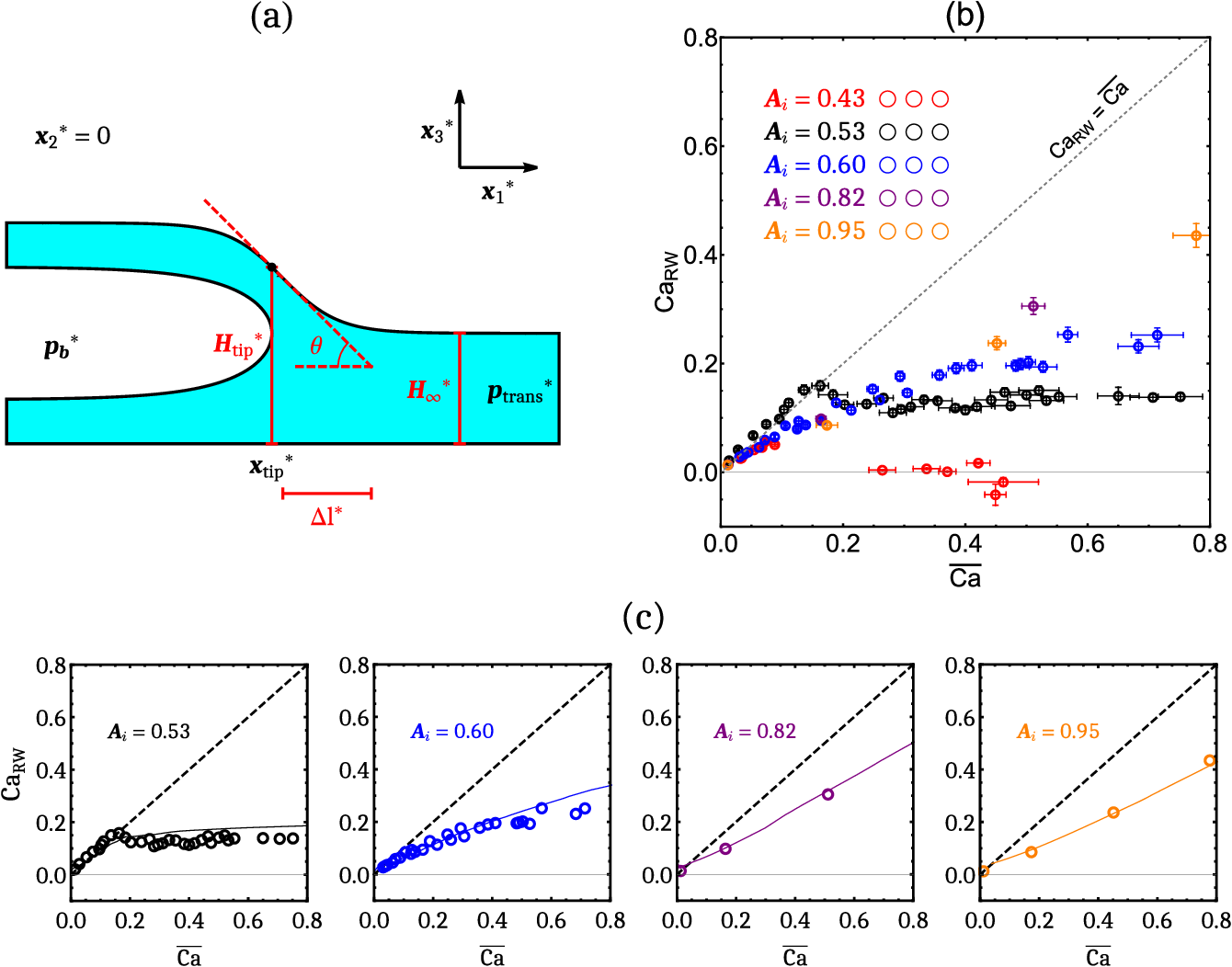}
	\caption{(a) Sketch of the fluid wedge in front of the finger. (b) Rigid-wedge capillary number (based on the pressure gradient in the fluid wedge ahead of the finger) as a function of the mean capillary number $\overline{Ca}$ of the propagating finger, time-averaged during the propagation over the ROI. Circles indicate experimental data with error bars denoting standard deviations within the ROI and the black dashed line corresponds to the limit $Ca_\mathrm{RW}=\overline{Ca}$. (c) Comparison between experimental data from (b) and numerical simulations shown with lines. The panels from left to right are for $A_{i}=0.53,\, 0.60,\, 0.82$ and $0.95$, respectively.}
	\label{Darcy_factor}
\end{figure}

We begin by examining the reopening region ahead of the finger. We define a flow metric which we use to establish the overall fidelity of the model by comparing experimental and numerical reopening flows across different $Ca$ and levels of initial collapse. When air is injected into a fluid-filled, collapsed elastic channel, the cross-sectional area of the collapsed channel expands and the fluid within it redistributes to accommodate finger propagation. This process takes place in a reopening region ahead of the finger tip where the local fluid-structure interaction flow sets the mode of finger propagation \citep{Callum2020, JuelARFM2018}. If we assume that the reopening is dominated by viscous stresses, then we can approximate the process as flow into a rigid wedge (RW). This geometry is illustrated in figure \ref{Darcy_factor}(a), which shows a schematic diagram of the vertical mid-plane of the channel ($x_2^*=0$) where the wedge-like reopening region has angle $\theta$, length $\Delta l^{*}$ and pressure drop $\Delta p^{*}$. At low $Ca$, flow into a compliant channel is closely captured by the flow into a rigid convergent laterally unbounded channel of pressure gradient $\Delta p^{*}/ \Delta l^{*}$, where the wedge angle $\theta$ is set by fluid-structure interaction \citep{Gaver1996,Jensen2002,PengLister2019}. This approximation is completely 2D, and since there is no in-plane curvature, only the transverse curvature contributes to the pressure jump across the interface. In this rigid wedge, the capillary number corresponding to the dimensionless tip speed is directly proportional to the pressure gradient across the wedge and, following the depth-averaged approach, can thus be defined as
\begin{equation}
Ca_\mathrm{RW} \equiv \frac{- b^{*2}(x_{\mathrm{1,tip}} , 0)}{12\gamma^{*}}\frac{\Delta p^{*}}{\Delta l^{*}}.
\label{Darcy_law_rescaled}
\end{equation}
Here, $b^{*}(x_{\mathrm{1,tip}} , 0) = b^{*}_{0}b(x_{\mathrm{1,tip}} , 0)$ is the dimensional height of the elastic sheet at the centreline of the channel and at the $x_{1}$ coordinate of the finger tip, the pressure drop is approximated by
\begin{equation}
\Delta p^{*} \approx (p_\mathrm{trans}^{*} - p_\mathrm{b}^{*}) + \frac{2\gamma^{*}}{ b^{*}(x_{\mathrm{1,tip}} , 0) },
\label{pressure_drop_darcy}
\end{equation}
where $p^*_\mathrm{b}$ is the air pressure in the finger and $p^*_\mathrm{trans}$ the pressure in the collapsed channel far ahead of the finger, and the length of the wedge $\Delta l^{*}$ can be estimated as
\begin{equation}
\Delta l^{*} = \frac{ b^{*}(x_\mathrm{1,tip},0) - b^{*}(\infty,0)}{\tan(\theta)},
\label{wedge_scale}
\end{equation}
where $b^{*}(\infty,0) = b^{*}_{0}b(\infty,0)$ is the dimensional height of the membrane at the centreline of the channel, beyond the wedge region (see figure~\ref{Darcy_factor}(a)). The geometric quantities $b^{*}(x_{\mathrm{1,tip}} , 0)$, $b^{*}(\infty,0)$ and $\theta$ were measured while the finger crossed the ROI, and remained constant even for unsteady finger propagation.

 Figure \ref{Darcy_factor}(b) shows a plot of $Ca_\mathrm{RW}$, which is estimated in our elastic channel based on the pressure gradient across the wedge, as a function of the actual capillary number $\overline{Ca}$ based on the finger speed. In the experiment, the finger speed $u_\mathrm{f}^{*}$ could exhibit measurable fluctuations depending on parameters and thus, we use its time-averaged value $\overline{u_\mathrm{f}^{*}}$, over the time interval during which the finger-tip propagates in the ROI, to calculate a time-averaged capillary number $\overline{Ca} = \mu^{*}\overline{u_\mathrm{f}^{*}}/\gamma^{*}$. For the smallest values of $\overline{Ca}$ investigated ($\overline{Ca}<0.1$), we find $Ca_\mathrm{RW} \simeq \overline{Ca}$ as expected in a laterally unbounded rigid wedge. For low levels of collapse $A_i \ge 0.8$, $Ca_\mathrm{RW}$ is approximately proportional to $\overline{Ca}$ over the range of $\overline{Ca}$ investigated. The reduced slope of the $Ca_\mathrm{RW}$ curve, compared to that of rigid wedge, is due to changes in the finger geometry in the tip region. The fact that there is linear relationship indicates that the fluid redistribution associated with reopening is primarily shaped by viscous and capillary forces, so we refer to this regime as viscous reopening. This viscous reopening regime also occurs for larger initial collapse, with the datasets for $A_i=0.60, \, 0.53$ and $0.43$ exhibiting quasi-linear behaviour as a function of $\overline{Ca}$ for small values of $\overline{Ca}$. 

As $\overline{Ca}$ increases, the gradient of the $Ca_\mathrm{RW}$
curve for $A_i=0.60$ progressively decreases. This trend is enhanced
for $A_i=0.53$, in which case $Ca_\mathrm{RW}$ becomes approximately
constant at a threshold value, $\overline{Ca} \approx 0.17$,
indicating saturation of the pressure gradient over the fluid wedge,
and thus of the viscous stresses, while the velocity of the finger
continues to increase. The pressure gradient saturates
because both the elastic wall and air finger have reached their
limiting geometric configurations. Therefore, the fluid wedge no longer
changes its shape and the pressure drop across it remains
constant on the viscous scale, as also found in three-dimensional
elastic tubes by \cite{HazelHeil2003} for sufficiently high
propagation speeds.
 In fact, for $A_i= 0.43$, $Ca_\mathrm{RW}$ drops to values close to zero for $\overline{Ca}>0.1$ (red points in figure~\ref{Darcy_factor}(b)), which suggests that the influence of viscous stresses becomes negligible for high levels of collapse approaching the point of opposite wall contact. This is because the elastic sheet steepens significantly in the reopening region, leaving it occupied mostly by the peeling finger. In this configuration, the value of $b^{*}(x_{\mathrm{1,tip}} , 0)$ becomes so small that the driving pressure difference $( p_\mathrm{trans}^{*} - p_\mathrm{b}^{*})$ is counteracted by a large capillary pressure $\frac{2\gamma^{*}}{b^{*}(x_{\mathrm{1,tip}} , 0)}$. The result is a small estimated pressure gradient $\frac{\Delta p^{*}}{\Delta l^{*}}$ and therefore, a small $Ca_\mathrm{RW}$. Thus, the finger is primarily shaped by elastic and capillary forces \citep{Ducloue2017a,Callum2020}, so we refer to this regime as elastic reopening.

The comparison between experiments and steady numerical simulations is detailed with individual plots of each level of collapse in figure \ref{Darcy_factor}(c). Since these are steady numerical simulations, $Ca$ is used as $\overline{Ca}$. A comparison is not shown for $A_{i} = 0.43$, because of difficulties in resolving the simulations for levels of collapse very close to opposite wall contact, when the fluid layer in the channel becomes very thin. However, the numerical model (solid lines) shows remarkable quantitative agreement with the experimental data (circles) for levels of collapse $0.53 \le A_i \le 0.95$. This indicates that our lubrication-based fluid-structure interaction model captures the key physics underlying finger propagation in the experiment. We find an initial-collapse-dependent threshold value of $\overline{Ca}$ that divides viscous reopening from elastic reopening. Viscous reopening occurs at low $\overline{Ca}$ and is the only observed behaviour at low levels of initial collapse. Elastic reopening occurs at high $\overline{Ca}$ when the initial collapse is sufficiently large. We now proceed to discuss the modes of finger propagation associated with these different reopening regimes.

\subsection{Viscous reopening in weakly collapsed channels}
\label{slightly_collapsed_paper_2}

\begin{figure}
\centering{
\includegraphics[scale=0.50]{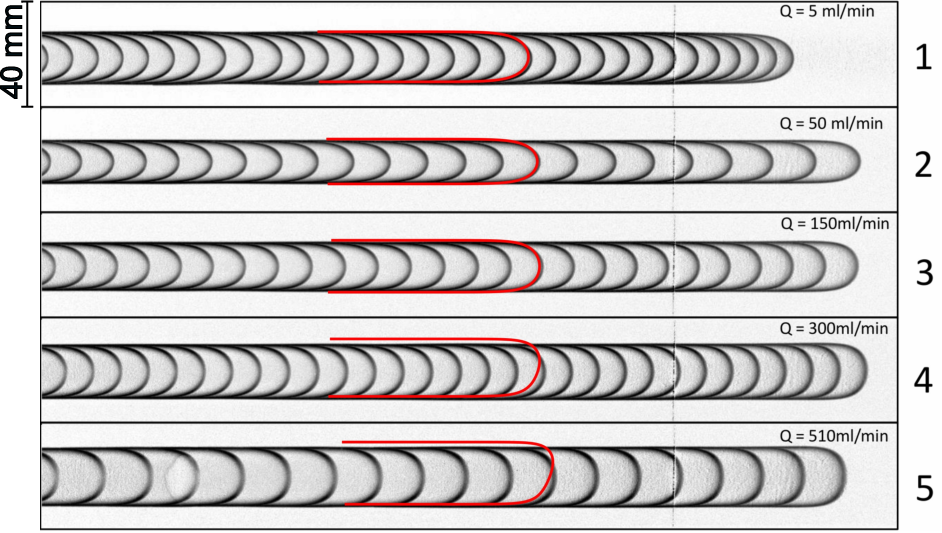}
\caption{Steadily propagating fingers for $A_{i}=0.95$. In experiments (1-5), the mean value of the capillary number over the ROI is $\overline{Ca} = 0.01, 0.17, 0.45, 0.78, 1.23$ respectively, while time intervals between the interfaces are $3.0$, $0.3$, $0.1$, $0.05$ and $0.05$ s, respectively. The interfaces in red are the steady numerical solutions for the same capillary number as in the experiments.}
\label{experimental_interfaces_095}
}
\end{figure}

For $A_{i} = 0.95$ and $A_{i} = 0.82$, all experiments gave steadily
propagating fingers, each of approximately constant capillary number, consistent with previous studies of airway reopening \citep{Grotberg2001, HazelHeilEtAl2012, Ducloue2017b}. Figure \ref{experimental_interfaces_095} shows sequences of time-lapse images from experiments for $A_{i} = 0.95$. Steady propagation is indicated by the fact that the shape of the finger does not vary as it propagates and that the consecutive interfaces are uniformly spaced. As the flow rate is increased from $Q^*=5$~ml/min (panel 1) to $50$~ml/min (panel 2), the symmetric finger narrows and its tip curvature increases. This trend is reversed upon further increase of the flow rate for $Q^* \ge 150$~ml/min (panels 3-5) with the finger widening and reducing its tip curvature. The red interfaces in each panel are finger profiles from steady numerical calculations with the same capillary number as the experiments. They accurately capture the finger profile in panels (1-3) but differences appear in panels (4-5), where the numerical solution is asymmetric about the centreline of the channel, in contrast to the experimental fingers which remain symmetric. 

\begin{figure}
 	\includegraphics[scale=0.75]{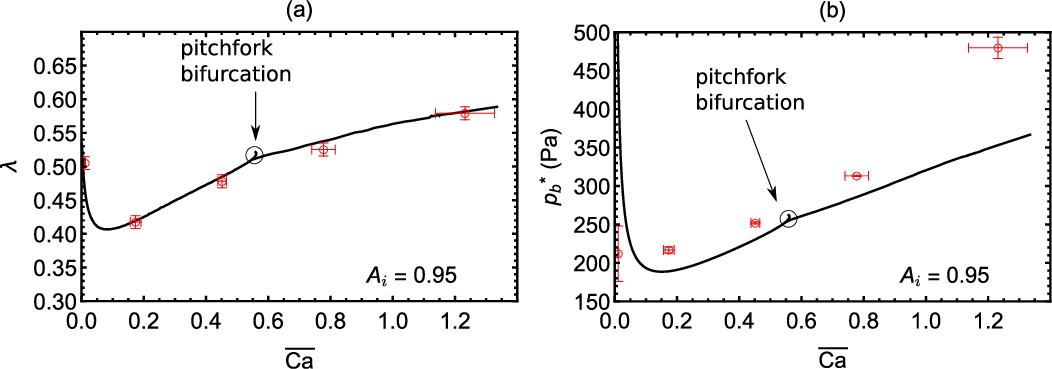}
	\caption{Slightly collapsed channel, $A_{i}=0.95$. Plots of (a) the finger width $\lambda$ and (b) the bubble pressure $p_\mathrm{b}^*$ as functions of the mean capillary number $\overline{Ca}$. The solid black lines depict the numerical steady solutions, while the red symbols indicate mean experimental values with error bars corresponding to standard deviations within the ROI. In both figures, we identify a pitchfork bifurcation where the symmetric round-tipped branch bifurcates into an asymmetric flat-tipped branch.}
	\label{Ainf_095_pushing_peeling}
\end{figure}

The non-monotonic variation of the finger width $\lambda$ as a function of $\overline{Ca}$ is plotted in figure \ref{Ainf_095_pushing_peeling}(a), with good agreement between experiments (red symbols) and steady simulations (black lines). The finger width only decreases with increasing $\overline{Ca}$ for the smallest values investigated, $\overline{Ca}<0.1$. This is the limit where $Ca_{RW}$ tends to $\overline{Ca}$ in \S \ref{darcys_factor_paper_2} and a decreasing finger width concurs with displacement flows in rigid Hele-Shaw channels \citep{Tabeling1986}.
The steady simulations undergo a pitchfork bifurcation near $\overline{Ca} = 0.55$, where symmetry about the centreline of the channel is lost. This bifurcation is responsible for the morphological change seen between panels (3) and (4) of figure~\ref{experimental_interfaces_095}.
In figure \ref{Ainf_095_pushing_peeling}(b), the steady simulations
(solid line) capture the overall increase of the experimental bubble
pressure (red circles) with increasing $\overline{Ca}$. The turning
point at $\overline{Ca} \simeq 0.15$ in the numerical curve is a
signature of compliant channel reopening, which indicates a transition
from pushing to peeling fingers
\citep{Gaver1996,HazelHeil2003,Callum2020}.
In the viscous pushing regime, the
  $p_{\mathrm{b}}^{*} \ vs \ \overline{Ca}$ curve has a negative
  slope and a large volume of fluid is displaced by the finger
  tip. In the viscous peeling regime,  the $p_{\mathrm{b}}^{*} \ vs
  \ \overline{Ca}$  slope becomes positive and very little fluid is
  pushed by the finger tip. In both these regimes, however,
  $Ca_\mathrm{RW}$ is proportional to $\overline{Ca}$, indicating
  the dominant balance is between viscous and capillary stresses.

In contrast to the numerical results, the experimental data does not
indicate a turning point at low $\overline{Ca}$, but the bubble
pressure fluctuates significantly for the smallest value of
$\overline{Ca}=0.01$, with variations of $\pm 20\%$ over the ROI,
despite an approximately constant velocity. This may be attributed to
the presence of gravity forces in the experiment
\citep{HazelGravity}. The increasing bubble pressure of peeling
fingers with increasing $\overline{Ca}$ means that the reopening
height of the channel increases and thus, volume conservation requires
the fluid behind the finger tip to occupy a smaller fraction of the
channel width \citep{Ducloue2017b}. Hence, the finger width $\lambda$
increases with $\overline{Ca}$ as shown in figure
\ref{Ainf_095_pushing_peeling}(a). We note that the variability of the
capillary number over the ROI increases with $\overline{Ca}$ (up to
$\pm 7\%$). This is consistent with thinner liquid films around the
peeling finger, which makes the finger more sensitive to channel
imperfections that can affect its velocity
\citep{Callum2020}. The experimental data point with
  the highest $\overline{Ca}$ in
  figure~\ref{Ainf_095_pushing_peeling}(b), is not in good agreement
  with the numerical prediction. We attribute this discrepancy to the
  fact that the experimental bubble pressure (500~$Pa$) is far outside
  the range of pressures $(-200~Pa < p <+200~Pa)$ for which the
  constitutive channel
  law used in our model was calibrated, see appendix
  \ref{appendix_a} for details of the calibration. The elastic sheets
  used in the experiments stiffen under inflation, meaning that in
  order to inflate the channel by the same amount higher
  pressures are required in the experiments than in the model; the
  level of channel inflation is set by conservation of mass. The
  stiffening increases nonlinearly, explaining why there is such a
  dramatic increase in the difference between experimental and
  numerical pressures at the highest value of $\overline{Ca}$.

At $A_{i} = 0.82$, we observe similar trends in both experiments and simulations; see Appendix \ref{appendix_c}. However, the turning point in the numerical simulations is displaced to a 40\% lower value of the capillary number. Moreover, a pitchfork bifurcation that leads to asymmetric fingers \citep{Fontana2020}, indicated in figure~\ref{Ainf_095_pushing_peeling} for $A_i =0.95$ by the emergence of a solution branch at $\overline{Ca}\simeq 0.55$, is displaced to smaller $\overline{Ca}$ for $A_i =0.82$ and a symmetry-breaking bifurcation is also observed in the experiments.

\subsection{Elastic reopening in strongly collapsed channels}
\label{multiple_modes_paper_2}

\begin{figure}
	\centering{
		\includegraphics[scale=0.75]{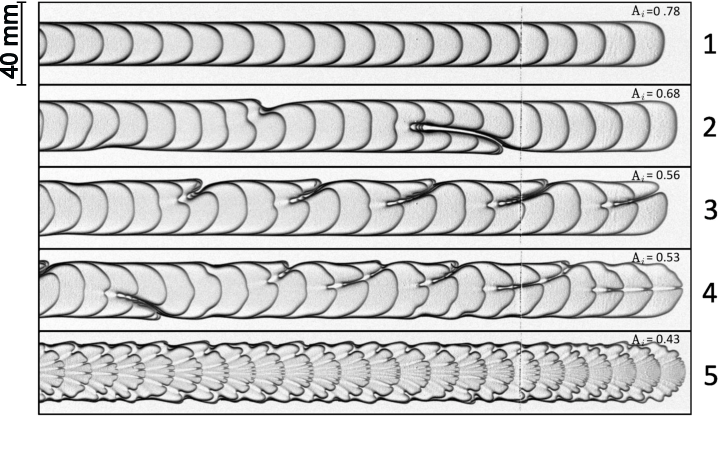}
	}
\caption{Experimental images of fingers propagating at a fixed mean capillary number $\overline{Ca}=0.36$ over the ROI. From top to bottom the initial level of collapsed increases. In experiments 1 to 5: $A_{i} = 0.78,~ 0.68,~ 0.56,~ 0.53,~ 0.43$. The time interval between interfaces is the same for all experiments, $0.14$~s.}
\label{Callum_Thesis_Fig3_12}
\end{figure}

We now turn to the elastic reopening regime which was identified in \S \ref{darcys_factor_paper_2} for increasing levels of collapse. Figure \ref{Callum_Thesis_Fig3_12} shows experimental sequences of time-lapse images of finger propagation at $Ca=0.36$, for five values of the initial level of collapse $0.43 \le A_i \le 0.78$. Steadily propagating fingers are only observed in the least collapsed channel with $A_i =0.78$ (panel 1), which reopens viscously based on the findings of \S \ref{darcys_factor_paper_2}. By contrast, finger propagation is unsteady for the larger levels of collapse (panels 2--5) where a dominantly elastic reopening regime is expected. The unsteadiness is associated with the appearance of interface dimples on the finger tip, which can grow into clefts and advect around the curved tip as the finger propagates. They arise either intermittently or approximately periodically. For the largest initial collapse ($A_i =0.43$), the interface indentations form small-scale corrugations on the finger tip which cause the characteristic feathered pattern previously highlighted by \cite{Callum2020}.

The disjoining pressure condition introduced in \S \ref{Model_paper_2} enables us to extend the numerical simulations benchmarked by \cite{Fontana2020} to capture unsteady finger propagation prevalent in the elastic reopening regime. We focus on unsteady finger propagation at $A_i=0.53$ and compare numerical findings with experiments in \S \ref{t_evo_unstable_modes_paper_2} . We then compute the underlying bifurcation structure in \S \ref{p_Ca_relations_paper_2} to gain insight into the transition from viscous to elastic reopening and unsteady propagation.

\subsubsection{Multiple modes of finger propagation}
\label{t_evo_unstable_modes_paper_2}

\begin{figure}
	\centering{
		\includegraphics[scale=0.48]{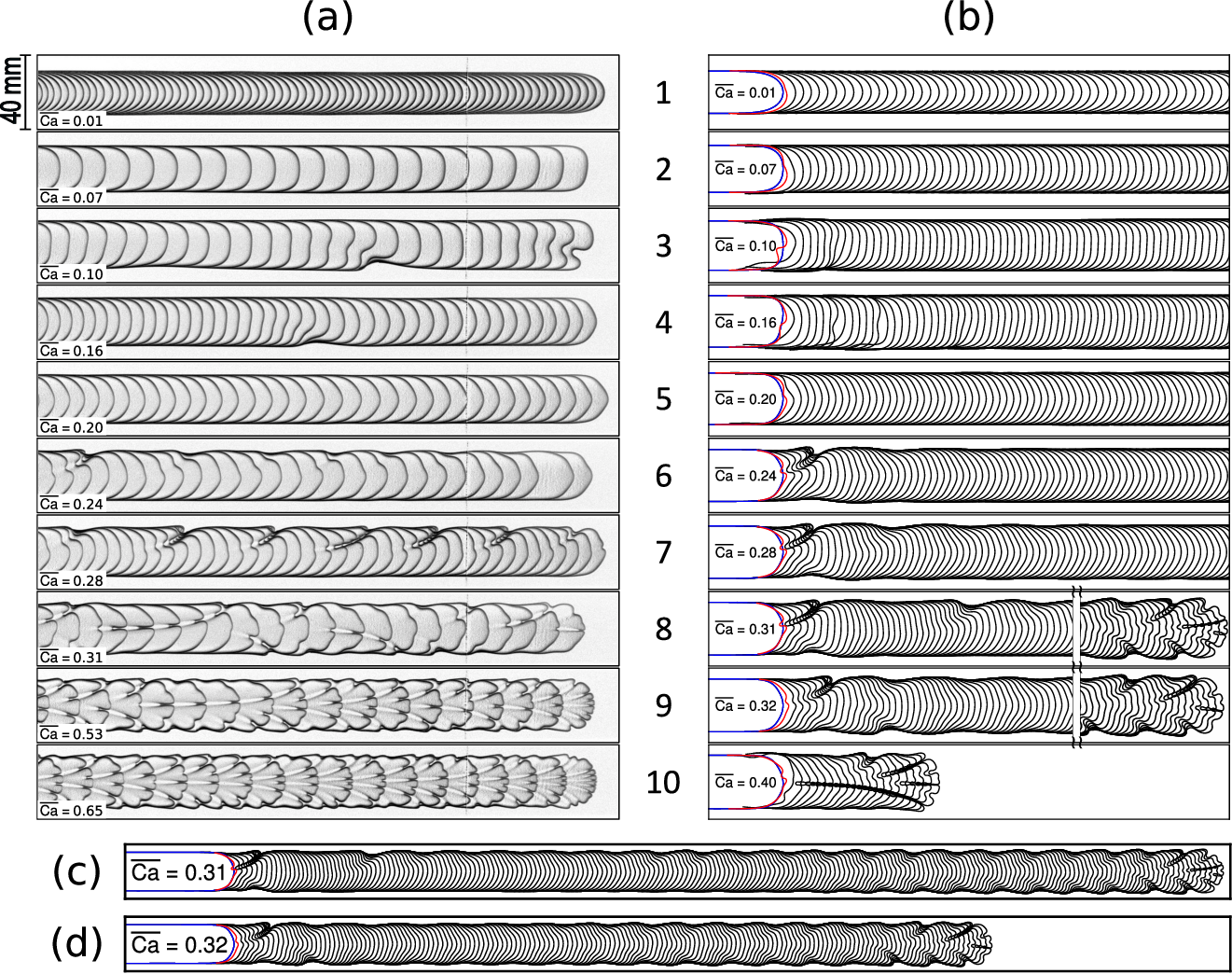}
		\caption{(a) Time evolution of experiments performed at a fixed $A_{i}=0.53$ and constant flow rate that increases from $5$~ml/min in experiment 1 up to $330$~ml/min in experiment 10. The time interval between the interfaces are, from 1 to 10: 1.17, 0.67, 0.40, 0.17, 0.17, 0.17, 0.17, 0.10, 0.17, 0.13 and 0.10~s, respectively. The third to last interface in experiment 6 is missing due to a camera fault. See also Supplementary Movies. (b) Time evolution of our time-dependent simulations at a fixed $A_{i}=0.53$ and constant flow rate that increases from $3$~ml/min in simulation 1 up to $160.3$~ml/min in simulation 10. The time interval between the interfaces is 0.08 in the non-dimensional scale, which results in 1.486, 0.205, 0.134, 0.078, 0.062, 0.052, 0.043, 0.038, 0.037, 0.028~s from 1 to 10, respectively. The initial steady solution in blue and the perturbed profile at $t_p$ in red are the first two finger profiles in each time-sequence. See also Supplementary Movies. (c-d) Complete domain of the time-simulations from panel 8 and 9 are depicted in (c) and (d), respectively.}
		\label{unstable_modes}
	}
\end{figure}

Figure \ref{unstable_modes}(a) shows the wide variety of finger
propagation observed experimentally for $A_i=0.53$ and flow rates
between $5$~ml/min in panel 1 and $330$~ml/min in panel 10 ($0.01 \le
\overline{Ca}  \le 0.65$). We classify the morphological structures seen in the experiments into four categories: (i) round-tipped, a finger propagating steadily with a symmetric curved front, depicted in panel 1 of figure~\ref{unstable_modes}(a); (ii) flat-tipped, a finger propagating steadily with an asymmetric flat front, depicted in panel 2 of figure~\ref{unstable_modes}(a); (iii) pointed, a finger propagating steadily with a symmetric triangular shaped front, depicted in panel 5 of figure~\ref{unstable_modes}(a); and (iv) feathered, a finger propagating unsteadily with small-scale perturbations continuously developing at the tip, depicted in panel 10 of figure~\ref{unstable_modes}(a).

For the lowest values of $\overline{Ca}$ (panels 1 and 2), channel reopening is associated with the steady propagation of round-tipped symmetric and flat-tipped asymmetric fingers similar to those discussed in weakly collapsed channels in \S \ref{slightly_collapsed_paper_2}. For $\overline{Ca}=0.10$ (panel 3), the asymmetric finger is prone to intermittent growth of tip perturbations, which is reminiscent of tip-splitting events in viscous fingering in rigid channels \citep{Couder2000}. The flat-tipped asymmetric state transitions to a pointed finger in panel 4 ($\overline{Ca}=0.16$). The appearance of the pointed finger, which is observed over the entire ROI in panel 5, coincides approximately with the saturation of $Ca_{RW}$ in figure \ref{Darcy_factor}, which marks the transition to the elastic reopening regime. These pointed fingers were not observed within the range of flow rates investigated for $A_i = 0.60$ (appendix \ref{appendix_c}). 

Panels 6--10 in figure \ref{unstable_modes}(a) all show unsteady finger propagation ($\overline{Ca} \ge 0.24$). In panel 6, dimples appear almost periodically on one side of the pointed finger tip and thus the finger propagation is asymmetric about the centreline of the channel. These periodic indentations grow into clefts in panel 7. By panel 8, the interface dimples form on both sides of the finger tip in alternation and the pattern of finger propagation appears mildly disordered. As discussed in \S \ref{slightly_collapsed_paper_2}, the finger width increases with increasing $\overline{Ca}$ thus reducing the curvature of the finger tip. This leads to feathered modes of propagation in panels 9 and 10 because the weakly curved front of the finger tip promotes the appearance of small stubby fingers which form a regular corrugation \citep{Ducloue2017a}. Variation of the characteristic width of these interfacial protrusions along the ROI could be due to imperfections in the channel but comparison between panels 9 and 10 suggests that their width decreases overall with increasing $\overline{Ca}$. The finger propagation is unsteady because the weakly-curved finger tip advects the small finger-like protrusions sideways from the centreline, thus broadening the central protrusions which in turn undergo tip-splitting. This results in a delicate cycle of small-scale stubby finger generation on the weakly curved front, where the length scale of the fingering decreases continuously as the finger propagates (panels 9, 10). Remarkably, the interface in these experiments is not prone to self-intersection despite its fine indentation, as discussed in \S \ref{disjoin_paper_2}. This was also the case with the previously mentioned clefts in panels 7 and 8, which always remained sufficiently short and broad.

Figure \ref{unstable_modes}(b) shows time simulations performed for the same values of mean capillary numbers as in the experiments in column (a), except for panels 9 and 10 where $\overline{Ca}$ is smaller then in the corresponding panels in column (a). We made the choice of taking smaller steps in $\overline{Ca}$ for the simulations, so we could capture the change in the transient behaviours that precedes the onset of the feathered mode. Time-simulations were initialised with steady solutions (blue contours) which were similar to the finger envelopes observed experimentally: round-tipped symmetric fingers (panels 1, 3), flat-tipped asymmetric fingers (panel 4) and pointed fingers (panels 5--10); see \S \ref{p_Ca_relations_paper_2}. The distorted finger shapes after perturbation of the steady solution using equation (\ref{pressure_perturbation}) are shown with red contours at $t=t_p=0.04$. In panels (1--7), steady modes of finger propagation are recovered over a distance of the order of a channel width. The transitions from symmetric round-tipped (panels 1, 2) to asymmetric flat-tipped (panels 3, 4) fingers, and to symmetric pointed fingers (panels 5--7), occur for similar values of $\overline{Ca}$ as in the experiment. The absence of finger indentations in the simulations in panels (1--5) means that the disjoining pressure did not affect the evolution of these fingers. In the simulations in panels (6--7), the initial perturbation evolved into an indentation but interface self-intersection at this indentation was prevented by the disjoining pressure, thus allowing the simulations to proceed.

The first instance of unsteady finger propagation occurs at $\overline{Ca}=0.31$ (panel 8) in the simulations and at $\overline{Ca}=0.24$ (panel 6) in the experiments. For larger respective values of $\overline{Ca}$, both experiments and time-simulations only support unsteady finger propagation. For sufficiently large values of $\overline{Ca}$, regardless of initial transient, the finger develops into the feathered mode.

Due to long transients, simulations in panels 8 and 9 only show the early and later stages of the finger propagation. Figures~\ref{unstable_modes}(c,d) show the full length of the numerical domain for these two panels. They indicate that the numerical model captures the unsteady dynamics observed in the experiment, where the pointed finger develops periodic perturbations (Figure~\ref{unstable_modes}(a), panels 6, 7) . In figure~\ref{unstable_modes}(c), asymmetric oscillations destabilize the symmetric pointed finger after a long transient. The perturbation develops a cleft close to the fingertip, which is advected and undergoes tip-splitting at approximately the same time. This process is repeated resulting in the feathered pattern. In figure~\ref{unstable_modes}(d), we observe a similar transition from pointed finger to feathered mode, but the transient oscillatory perturbation is symmetric instead. This transition from asymmetric to symmetric perturbations is seen in the experiments (Figure~\ref{unstable_modes}(a), panels 7, 8) with similar values of $\overline{Ca}$. However, the finite length of the experimental channel means that we cannot say for certain if the patterns seen in panels 7 and 8 are long term transients or periodic states.

Finally, at $\overline{Ca}=0.40$ (figure \ref{unstable_modes}(b), panel 10) the simulation depicts a finger that evolves from the symmetric pointed to the feathered state after a very brief transient. The time-dependent simulations of the feathered modes are interrupted when the width of the small-scale finger becomes comparable to the channel height $b_{0}^{*}$. At this point the depth-averaged model is no longer reliable. As in the experiments, the typical lengthscale of the small-scale fingering seen in the simulations decreases as $\overline{Ca}$ increases. However, the simulations feature small-scale indentations that are consistently deeper than in the experiments. Given that the typical indentation width is comparable to the unperturbed channel height $b_{0}^{*}$, there will be three-dimensional effects in the experiments that cannot be captured by our model. At this stage, regardless of the calibration or the choice of functional form for the disjoining pressure, three-dimensional Stokes simulations are necessary to correctly capture the interfacial interaction between the small-scale fingers.

\begin{figure}
\centering{
\includegraphics[width=\textwidth]{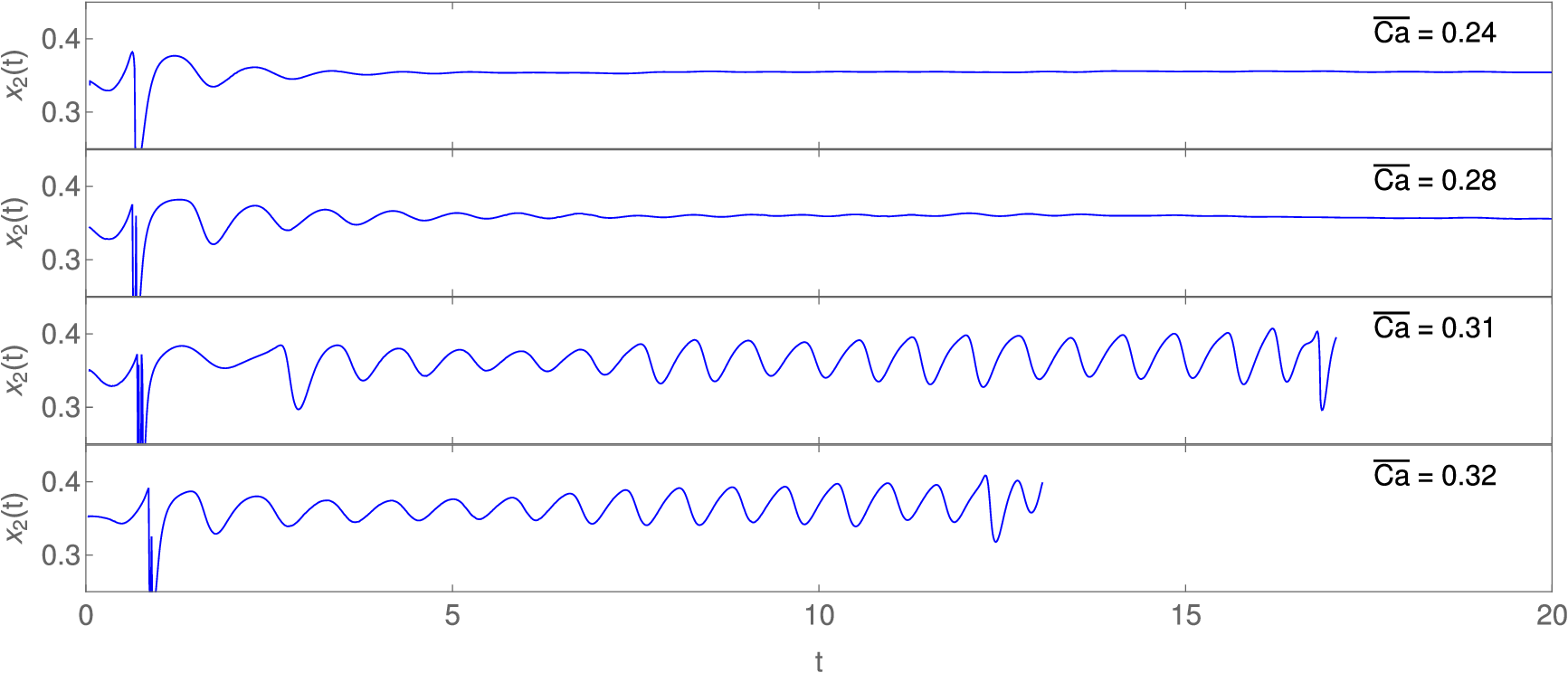}
\caption{Time evolution of the $x_{2}$ coordinate of a point on the interface at a fixed axial distance $dx=0.5$ behind the finger tip. For (a) $Ca=0.24$, (b) $Ca=0.28$, (c) $Ca=0.31$ and (d) $Ca=0.32$. These values are extracted from the simulations depicted in figure~\ref{unstable_modes}(b).}
\label{Feathered_periodicity}
}
\end{figure}

The propagation of the experimental feathered modes gives no indication that a periodic state is being reached, instead we can see that the length scale of the fingering pattern is always refining for the entire length of the channel. Similarly, the simulations of the feathered mode stop once the length scale of the fingering pattern reaches the length scale of the channel height. In figure~\ref{Feathered_periodicity} we gauge the apparent periodicity of the feathered modes by tracking the time evolution of the $x_{2}$ component of a point $\Pi = (x_{\mathrm{1,tip}} - dx , x_{\mathrm{2}}(t))$ on the finger interface at a fixed axial distance $dx=0.5$ behind the finger tip $\Pi_\mathrm{tip} =  (x_{\mathrm{1,tip}} , x_{\mathrm{2,tip}})$. For both $Ca=0.24$ and $Ca=0.28$, after a transient due to the advected perturbations, $x_{2}$ component of $\Pi$ eventually reaches a steady value. For the feathered modes, figures~\ref{Feathered_periodicity}(c)-(d), although the oscillations persists as long as the simulations last, they never reach a periodic state. This gives us an indication that the simulations point to the same result seen in the experiments, that there is no periodic state being reached.

\subsubsection{Pressure-$Ca$ relations and bifurcation structure}
\label{p_Ca_relations_paper_2}

\begin{figure}
\centering{
\includegraphics[scale=0.45]{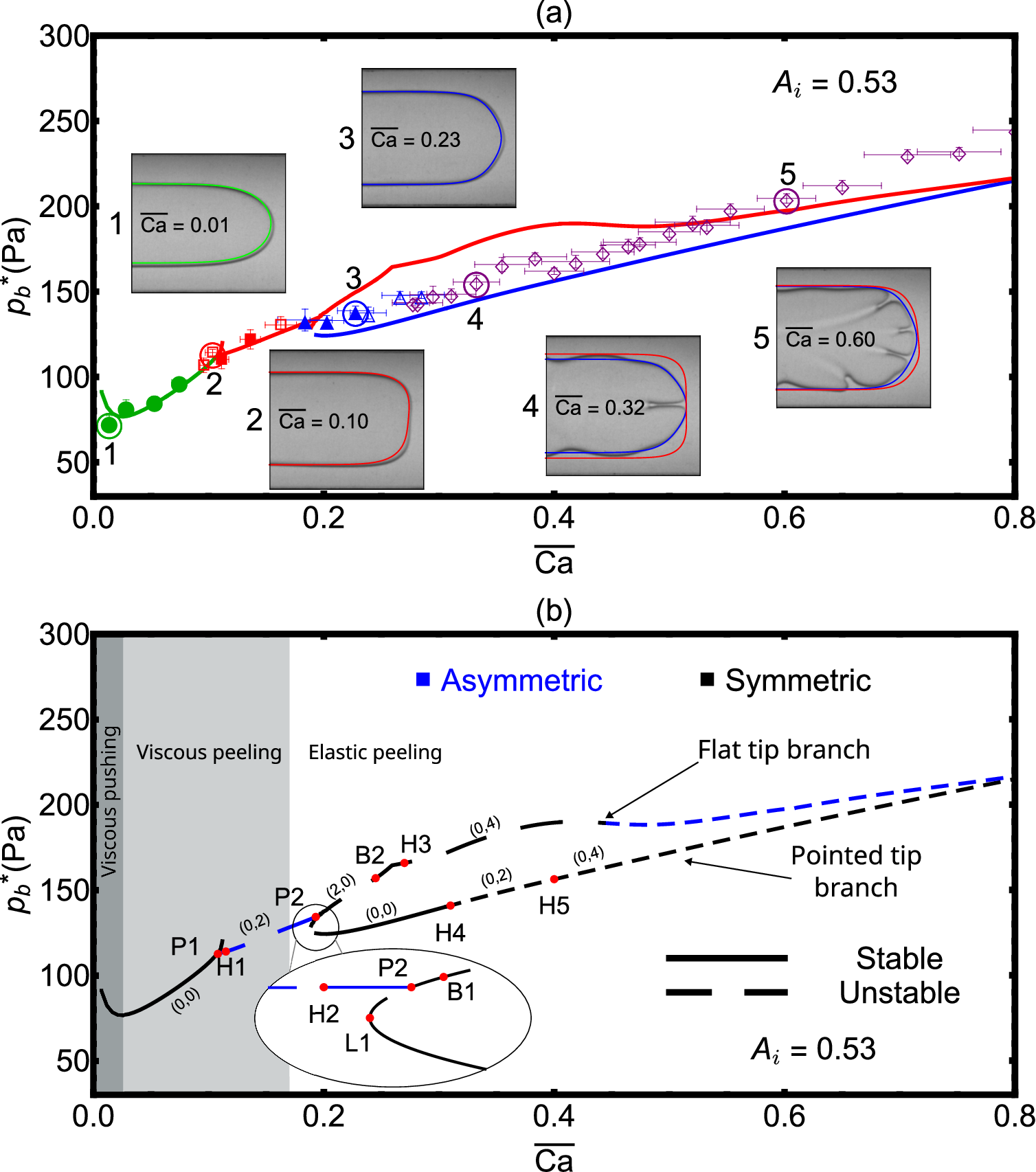}
\caption{Plot of the bubble pressure $p_\mathrm{b}^{*}$ as a function of the mean capillary number $\overline{Ca}$ at the fixed level of collapse $A_{i}=0.53$. (a) Experimental fingers are plotted as symbols. Green circles represent round-tipped, red squares represent flat-tipped (symmetric or asymmetric), blue triangles represent pointed-tipped and purple diamonds represent feathered fingers. Filled symbols represent steadily propagating experimental fingers, while empty symbols represent unsteady ones. Error bars indicate standard deviations of the capillary number within the ROI. Steady numerical solutions are presented as solid lines following the same color code from the experiments. Insets are experimental images with steady numerical interfaces at the same $\overline{Ca}$ superposed. They correspond to the numbered points in the graph. (b) Bifurcation diagram of the steady numerical solutions. Solid (dashed) lines represent stable (unstable) solutions. Black (blue) lines represent symmetric (asymmetric) solutions. 
The number of positive real eigenvalues $m$ and complex eigenvalues with positive real parts $n$, corresponding to instabilities, are indicated in brackets $(m,n)$ for each solution branch.
The relevant bifurcations are marked with red dots. $P_1$ and $P_2$ are pitchfork bifurcations, $H_1$-$H_5$ are Hopf bifurcations, $L_1$ is a limit point and $B1,\, B2$ indicate more complex transitions. The bifurcation structure around $P_2$ could not be fully detailed but findings are consistent with the picture provided in the inset.}
\label{p_Ca_A053}
}
\end{figure}

As previously reported by \cite{Fontana2020}, the model system exhibits a complex solution structure, with multiple co-existing steadily propagating states and numerous bifurcations.
Figure \ref{p_Ca_A053}(a) compares bubble pressure $p_\mathrm{b}^{*}$ in the experiments and steady numerical simulations as a function of the mean capillary number $\overline{Ca}$ for $A_{i} = 0.53$. As for weakly collapsed channels in \S \ref{slightly_collapsed_paper_2}, $p_\mathrm{b}^{*}$ increases with $\overline{Ca}$ indicating peeling fingers. The turning point in the pressure that indicates a transition to pushing in the numerical simulations is at the lower end of the capillary number range investigated ($Ca \simeq 0.03$). The symbols of different colours distinguish the different finger types observed experimentally. As shown in figure \ref{unstable_modes}, the steadily propagating round-tipped symmetric finger (solid green circle), flat-tipped asymmetric finger (solid red square) and pointed symmetric finger (solid blue triangle) occur in succession as $\overline{Ca}$ increases. The steady numerical simulations capture their bubble pressures and $Ca$ ranges quantitatively. The insets show examples of experimental fingers overlaid with a steady numerical solution at the same $\overline{Ca}$. In insets 1, 2, and 3 where the finger propagation is steady, there is excellent agreement between simulations and experiments for finger width and morphology. The unsteady experimental fingers discussed in \S \ref{t_evo_unstable_modes_paper_2}, which occur in the vicinity of transitions between modes and for $\overline{Ca} > 0.27$, are shown with open symbols. In the numerical simulations the solution branches for flat-tipped fingers (red), which include both symmetric and asymmetric solutions, and pointed symmetric fingers (blue) continue to values of the capillary number where only feathered fingers are observed in the experiment ($\overline{Ca} > 0.27$). For $\overline{Ca}$ up to $0.5$, the bubble pressure of the experimental feathered modes lies between the red and blue curves but closer to the latter. As $Ca$ increases further, the shape and bubble pressure of the flat-tipped finger in the simulations approach those of pointed fingers, with pressure below the experimental data. The insets indicate that the overall finger shape upon which instabilities develop in the experiment is the pointed finger (blue) rather than the flat-tipped finger (red); see inset 4. As $\overline{Ca}$ increases, it becomes progressively more difficult to distinguish these two modes; see inset 5.

In figure~\ref{p_Ca_A053}(b) we highlight the symmetry and stability of the steady solutions which were computed by imposing fixed $Q^{*}$ and $A_{i}$ to mimic the experiments. The different regimes are highlighted with shaded regions that are defined by the pushing/peeling transition, where the $P_b \ vs \ \overline{Ca}$ slope changes sign, and the viscous/elastic transition, estimated from ~\ref{Darcy_factor}(c). The solution structure is complex, consistent with that previously reported by \cite{Fontana2020}, but does not always have a direct connection to the observed time-dependent behaviour. Consequently, we have not pursued a detailed study of this bifurcation diagram, other than to confirm that all bifurcations and changes of stability are consistent with standard theory. There are, however, connections that can be made between the bifurcation diagram and experimental observations.

The round-tipped symmetric finger observed at low $\overline{Ca}$ loses stability to a pair of emerging asymmetric fingers at a supercritical pitchfork bifurcation ($P_1$), which is consistent with the experimental transition to a flat-tipped asymmetric finger at $Ca=0.1$. The asymmetric fingers are superposed in this projection because they have the same bubble pressure. The unstable symmetric branch emanating from ($P_1$) is double-tipped, which provides a mechanism for intermittent tip-splitting near $\overline{Ca}=0.1$ as seen in figure \ref{unstable_modes}, panel 3. The finger may visit the vicinity of this unstable branch in the experiment intermittently due to background perturbations (figure \ref{unstable_modes}(a), panel 3) or once following initial perturbation in the numerical system (figure \ref{unstable_modes}(b), panel 3). 

The pitchfork bifurcation $P_2$ occurs at $\overline{Ca}=0.18$ and the
bifurcation scenario in its vicinity is consistent with the
experimental transition from asymmetric flat-tipped fingers to pointed
fingers, which accompanies the abrupt saturation of $Ca_{\mathrm{RW}}$
in \S \ref{darcys_factor_paper_2}. This signals elastic reopening,
where $p_{\mathrm{b}}^{*}$ is proportional to $\overline{Ca}$, and thus suggests that finger morphologies such as pointed and feathered modes are a feature of elastic vessel reopening \citep{HeapJuel2008, Callum2020, Ducloue2017b}. In the numerical model, the feathered modes emerge beyond the Hopf bifurcation $H_4$. Between $H_4$ and a subsequent Hopf bifurcation, $H_5$, the pointed finger has one pair of unstable complex conjugate eigenvalues, while after $H_5$ it has two pairs. However, in both regions, the frequencies of the eigenmodes do not match the typical oscillatory time scales of the feathered modes; see Figure \ref{Feathered_periodicity}. This suggests that these complex modes of propagation, discussed in \S \ref{t_evo_unstable_modes_paper_2}, correspond to fully nonlinear dynamics. It is worth pointing that there are likely to be other solutions, in addition to those presented here. For instance, \cite{Fontana2020} have found multi-tipped Romero--Vanden-Broek~\citep{Romero1982,VandenBroeck1983,mccue2015,green2017effect} solutions, calculated originally for the rigid channel. These solutions were not stabilized by the presence of the elastic sheet.

\section{Conclusion}
\label{Conclusion_paper_2}

We have explored the transition to complex pattern formation associated with two-phase flows in a Hele-Shaw channel, where the upper rigid boundary is replaced by an elastic sheet. This elastic upper boundary enables the collapse of the channel so that it adopts a prescribed non-uniform cross-sectional depth distribution and its cross-sectional area is reduced relative to the undeformed rectangular channel. Injection of air at constant flow rate into this initially-collapsed, liquid-filled channel leads to the propagation of an air finger, which reopens the channel by redistributing resident fluid ahead of its tip. We find experimentally that these propagating fingers exhibit unsteady dynamics for sufficient large level of initial channel collapse (small $A_i$) or large finger capillary number ($Ca$), and that the characteristic feathered modes first identified by \cite{Callum2020} are supported across a wide range of $A_i$ and $Ca$. 

To capture these complex finger patterns with numerical simulations, we extend the depth-averaged model developed by \cite{Fontana2020} to include an artificial disjoining pressure term. This term is added to circumvent self-intersection of the air finger interface which prematurely terminates the simulations, and occurs when the applied local interface perturbation grows into a cleft that is sufficiently deep and narrow. The fact that self-intersection of the air-finger interface is not observed in the experiments suggests that its presence in simulations is likely a result of the approximations made when simplifying the three-dimensional liquid-film dynamics. 

Although the inclusion of fluid film effects in the depth-averaged model is fundamental for both qualitative and quantitative agreement with experiments \citep{Fontana2020}, we find that the results from numerical simulations are unaffected by our precise choice of film model, i.e., whether the film thickness is set at the finger tip and thus has a uniform thickness \citep{PengPihler2015}, or varies with the local height of the finger \citep{Fontana2020}. This is because the choice of fluid film model only affects the system via mass conservation, so our results are unchanged provided we present them as a function of the capillary number rather than the imposed flow rate.

We started by quantifying the relative importance of viscous and elastic effects on the flow in the reopening region ahead of the finger tip. For this we compared the elastic system with the propagation of a finger into a rigid wedge \citep{PengLister2019,JuelARFM2018}. A rigid-wedge capillary number ($Ca_\mathrm{RW}$) was estimated by converting the pressure gradient ahead of the finger tip, using the depth-averaged theory, into a dimensionless finger speed set by viscous and capillary forces. We used this metric to establish the fidelity of the simulations across the range of $A_i$ and $Ca$ studied, by demonstrating notable quantitative agreement between numerical and experimental values of $Ca_\mathrm{RW}$. Comparing $Ca_\mathrm{RW}$ to $Ca$, which is measured based on the finger speed, we find an abrupt saturation of $Ca_\mathrm{RW}$ at a threshold value of $Ca$ for sufficiently large initial collapse. This saturation of viscous dissipation signals a transition from reopening dominated by viscous and capillary forces to dominant elastic and capillary effects. We refer to these distinct reopening regimes as `viscous' and `elastic', respectively. We find that the exotic finger morphologies observed experimentally, such as pointed fingers and the feathered mode, are associated with an elastic regime of reopening. 

A combination of experiments and numerical simulations was next used to explore the modes of finger propagation associated with the viscous and elastic regimes. In the viscous regime which corresponds to modest channel collapse and/or moderate values of $Ca$, we recover the non-monotonic variation of bubble pressure (and width) with $Ca$, which is characteristic of benchtop models of pulmonary airway reopening \cite{Gaver1996,Jensen2002}, and fundamentally distinguishes elastic-channel finger propagation from two-phase displacement flows in rigid Hele-Shaw channels \citep{SaffmanTaylor1958}.

The introduction of a disjoining pressure into the numerical model enables the study of the elastic regime, which occurs for increased channel collapse and/or $Ca$. This is because the disjoining pressure prevents the self-intersection of the interface when the imposed perturbation evolves into a cleft, e.g., for $Ca \ge 0.24$ when $A_{i}=0.53$. Remarkably, the extended numerical model captures the destabilization of the finger into feathered modes of propagation, and this process helps to shed light on the rich variety of pattern formation observed experimentally. The feathered modes of propagation emerge following long asymmetric ($Ca=0.31$) or symmetric ($Ca=0.32$) oscillatory transients, which are associated with patterns reminiscent of the oscillatory fingers observed in the experiment for $Ca=0.24$ and $0.28$, respectively. Furthermore, both experiments and simulations indicate that the feathered modes themselves may be transient. Increasingly fine indentation of the finger tip as it propagates indicates that the time-scale for tip-splitting at the interface is shorter than the time-scale for advection of the interface perturbation, and thus leads to ever refining fingering. These findings suggest that the unsteady patterns observed experimentally, characteristic of the elastic regime, may be continually evolving, with no evidence that these disordered modes converge to steady or periodic states. 

Finally, we find excellent quantitative agreement between experiments and stable steady numerical solutions. Steady solutions also match the bounding envelopes of unsteady fingers obtained in the experiments and time simulations. This means that the steady bifurcation structure calculated numerically is sufficient to predict the bubble pressure. However, the steady bifurcation diagram does not inform unsteady finger propagation prevalent in the elastic regime, in that time-dependent modes which bifurcate from the steady solutions have different characteristic time scales from the feathered modes observed in the experiments and time simulations. Hence, in spite of the apparent simplicity of this system, the dynamics observed in the experiments and time simulations reflect complex nonlinear dynamics that require finite amplitude perturbations of the steady bifurcation structure to be attained.

\section*{Acknowledgements}

This work was supported via EPSRC grants EP/J007927/1, EP/P026044/1 and EP/R045364/1. The preliminary model development was supported by the Leverhulme Trust under grant number RPG-2014-081.

\section*{Declaration of interests}

The authors report no conflict of interest.

\appendix

\section{Estimation of the pre-stress}
\label{appendix_a}

\begin{figure}
\centering{
\includegraphics[scale=0.25]{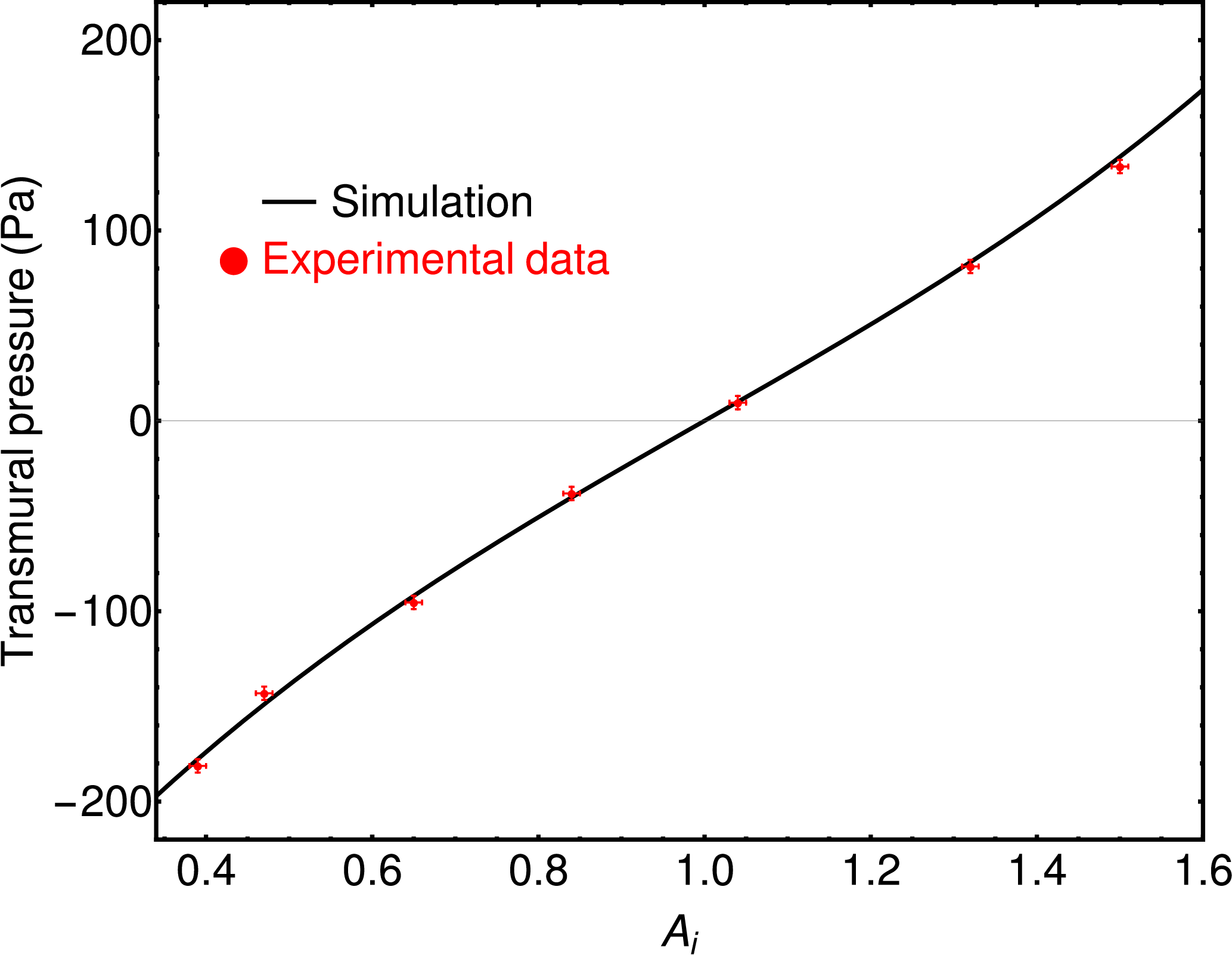}
\caption{Relationship between the transmural pressure and the level of collapse of the channel, $A_{i}$. The red circles indicate the experimental data and the solid black line corresponds to the numerical simulation for $\sigma_{22}^{(0)*}=32$~kPa.
}
\label{channel_law}
}
\end{figure}

We followed the methodology developed by \cite{Fontana2020} and used the pre-stress as a fitting parameter chosen to achieve the best quantitative agreement between experimental and numerical channel law, i.e., the constitutive relation between transmural pressure (difference between the pressure inside the channel and atmospheric pressure) and the level of collapse, $A_i$, in a channel filled with fluid at rest. The channel law was measured experimentally and modelled using the F{\"o}ppl--von K\'{a}rm\'{a}n equation to predict the shape of the elastic upper boundary of the channel under prescribed transmural pressure. Results are shown in figure \ref{channel_law}, where the red symbols show the experimental data, and the solid black line is the best fit computed by our model. The resulting pre-stress fitted by this procedure is $\sigma_{22}^{(0)*}=32$~kPa, $\sigma_{11}^{(0)*}=0$ and $\sigma_{12}^{(0)*}=0$.

\section{Uniform and non-uniform film thickness}
\label{appendix_b}

In \S \ref{Model_paper_2} we describe the model of fluid films developed by \cite{PengPihler2015} in which the thickness of the film between the finger and the channel boundary is a fraction $f_{1}(Ca)$ of the channel height, and modifies its transverse curvature by a factor of $f_{2}(Ca)$. The expressions for the functions $f_{1}(Ca)$ and $f_{2}(Ca)$ were derived assuming a rigid channel, but \cite{PengPihler2015} and \cite{Fontana2020} showed that this fluid film model yields excellent agreement with experiments when applied to a radial elastic cell and a rectangular elasto-rigid channel, respectively.

Here, we follow \cite{Fontana2020} and assume that the non-uniform thickness of the fluid film at any point $(x_{1},x_{2})$ on the finger is given by $f_{1}(Ca)b(x_{1},x_{2})$. Alternatively, as was done by \cite{PengPihler2015}, we could assume that the fluid film is generated at the finger tip, where the height of the channel is $b_{\mathrm{tip}}$. This yields a uniform film thickness $f_{1}(Ca)b_\mathrm{tip}$ across the finger region. Both choices correspond to approximations of the actual fluid-film distribution, which could be determined by a full 3D Navier-Stokes simulation. We also assume that the film model is a function of the global capillary number $Ca$ based on the averaged finger tip velocity, instead of a local capillary number based on the velocity at each point on the interface. Global and local capillary numbers take the same values in the case of steadily propagating fingers. For unsteady fingers, however, the velocity of the interface can vary significantly across short length-scales, particularly in the presence of small-scale fingering, but panels 8, 9 and 10 of figure \ref{unstable_modes}(b) suggest that our model is able to capture the feathered mode, which is the most complex mode of finger propagation observed in our system.

\begin{figure}
\centering{
\includegraphics[scale=0.8]{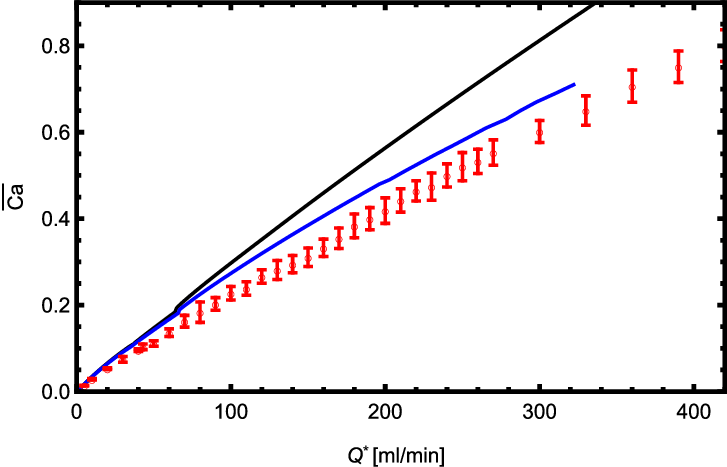}
\caption{Mean capillary number as a function of flow rate, for an initial level of collapse of $A_{i}=0.53$. The blue and black lines represent the numerical model with uniform and non-uniform film thickness distributions, respectively. The red symbols indicate the experimental data, where the error bars are the standard deviations of $\overline{Ca}$ over the ROI.}
\label{flow_x_Ca}
}
\end{figure}

\begin{figure}
\includegraphics[scale=0.5]{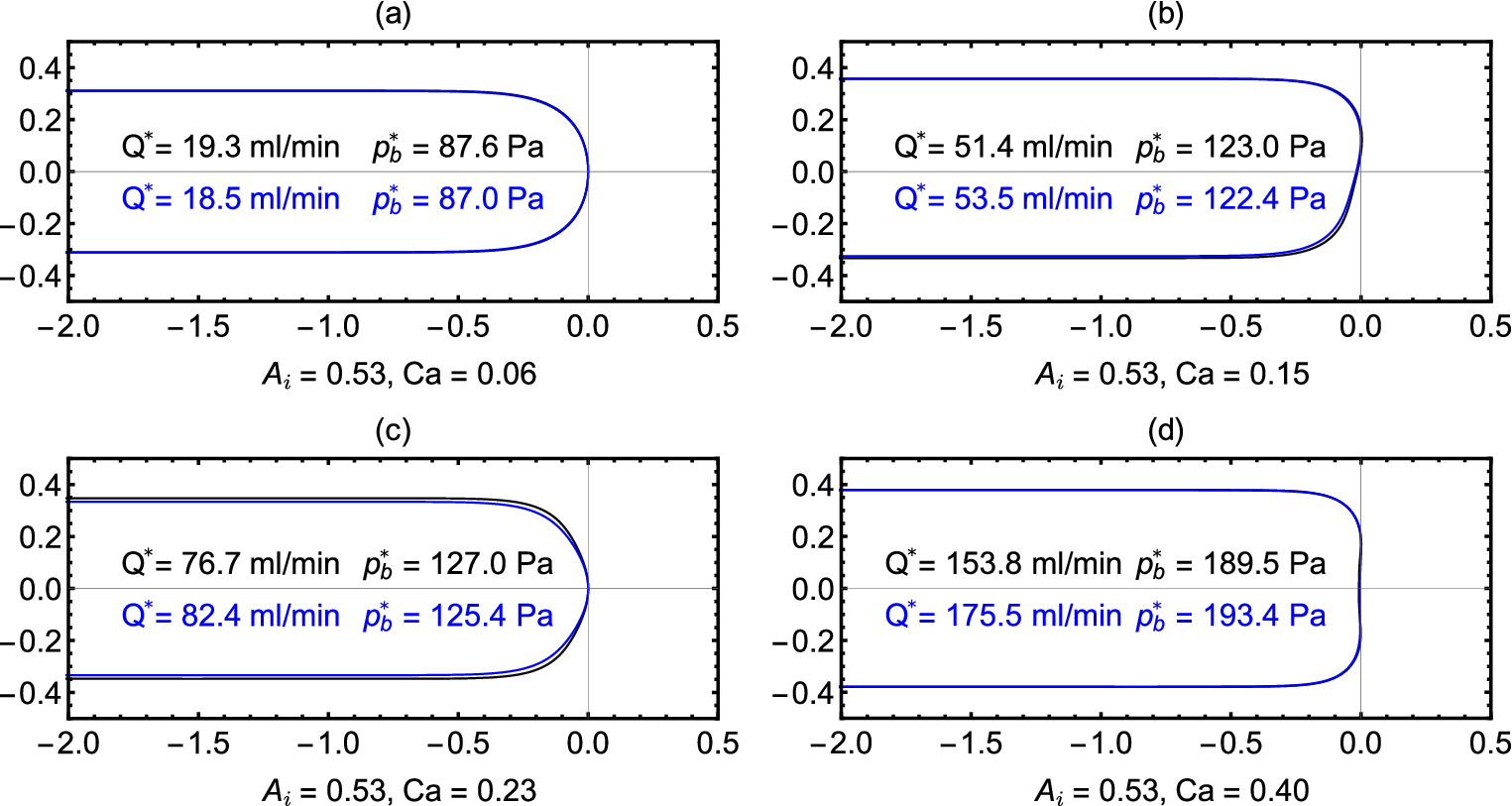}
\caption{Interfaces of steady solutions computed using the uniform (non-uniform) approach of the fluid film model plotted as the blue (black) solid line. The solutions are computed by fixing the value of capillary number: (a) $Ca = 0.06$,(b) $Ca = 0.15$,(c) $Ca = 0.23$ and (d) $Ca =40 $.}
\label{old_x_new_interfaces}
\end{figure}

The choice of either uniform or non-uniform films affects the finger cross-section and influences the relation between flow rate and finger speed, because of mass conservation. In figure \ref{flow_x_Ca} we compare the relationship between capillary number and flow rate in experiments and models with uniform and non-uniform films.  Although neither model quite matches the experimental data, the closest agreement is obtained with a uniform film thickness distribution. Figure \ref{old_x_new_interfaces} shows that the steady finger solutions obtained using the two variations of the film model superimpose for the same capillary number. Steady results from both models, such as finger shape, bubble pressure and transition between modes of propagation, are in remarkable agreement when we use the capillary number as the control parameter. Hence, throughout this manuscript, we present all experimental and numerical results in terms of capillary number. However, since the capillary number is time-dependent, in the in the unsteady propagation scenario, the non-uniform film thickness model is the natural choice for time-dependent simulations. Therefore, all numerical results presented in this manuscript, apart from those in this appendix, are obtained with a choice of non-uniform film thickness.

\section{Experimental data at $A_{i} = 0.82$ and $A_{i} = 0.60$}
\label{appendix_c}

\begin{figure}
\includegraphics[scale=0.5]{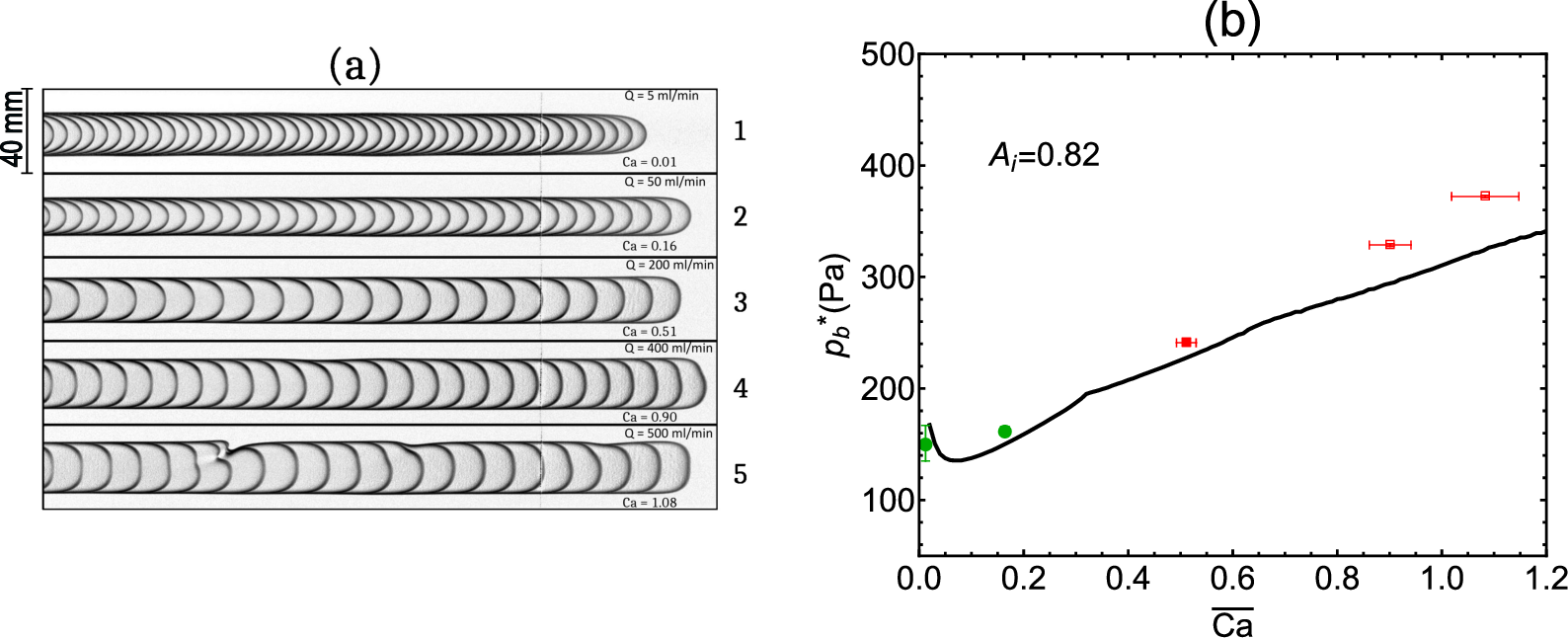}
\caption{(a) Time evolution of experiments performed at a fixed $A_{i} = 0.82$ and constant flow rate $Q^{*}$. From experiment 1 to 5, $Q^{*}$ increases from $5$ to $500$~ml/min. The constant time interval between the interfaces in each experiment is, from 1 to 5, $2.0$, $0.2$, $0.1$, $0.05$, $0.05$~s (b) Plot of bubble pressure $p_\mathrm{b}^{*}$ as a function of the mean capillary number $\overline{Ca}$ for experiments at $A_{i} = 0.82$. The stability and morphology of the fingers are presented using the same color/shape of experimental data points as in figure \ref{p_Ca_A053}(a). Steady simulations are presented as solid black lines.}
\label{Pb_Ca_Ainf_082_appendix_C}
\end{figure}

\begin{figure}
\includegraphics[scale=0.5]{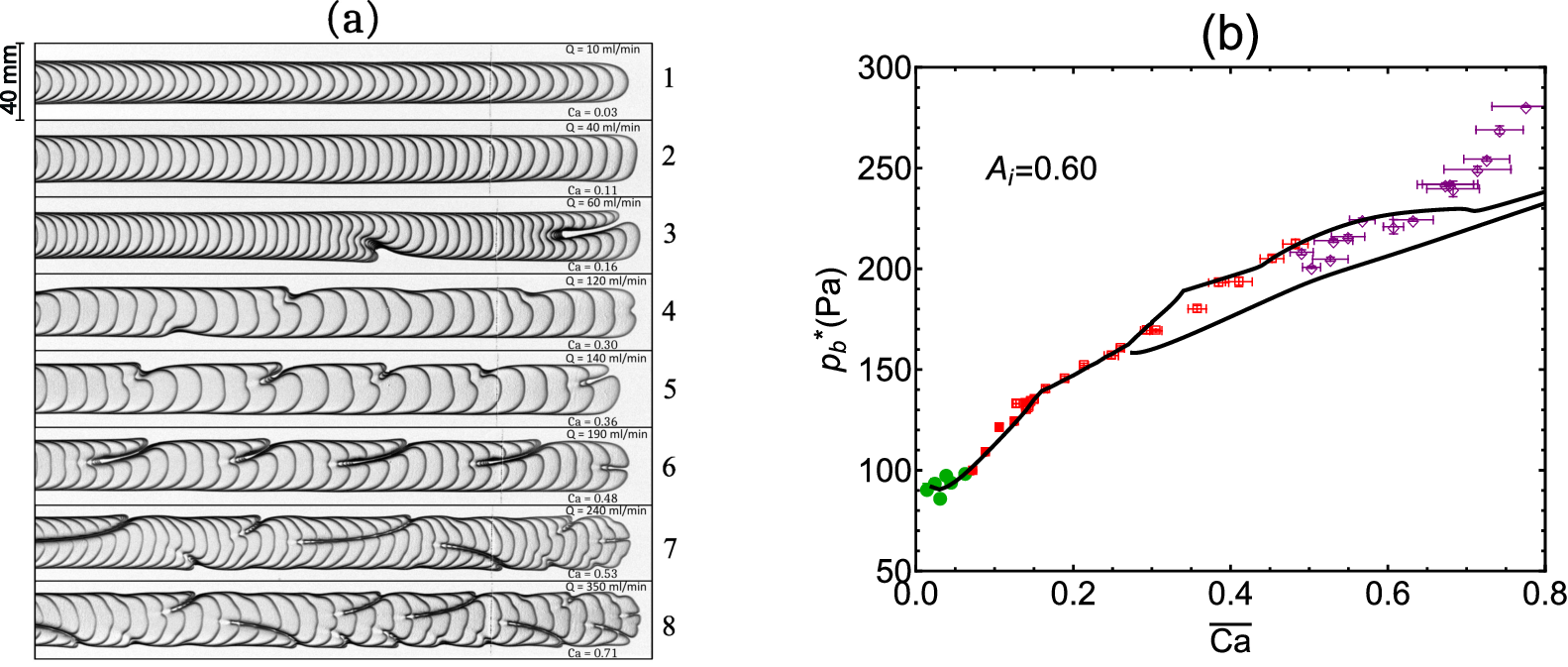}
\caption{(a) Time evolution of experiments performed at a fixed $A_{i} = 0.60$ and constant flow rate $Q^{*}$. From experiment 1 to 8, $Q^{*}$ increases from $10$ to $350$~ml/min. The constant time interval between the interfaces in each experiment is, from 1 to 5, $0.86$, $0.25$, $0.13$, $0.13$, $0.13$ $0.07$, $0.05$, $0.05$~s (b) Plot of bubble pressure $p_\mathrm{b}^{*}$ as a function of the mean capillary number $\overline{Ca}$ for experiments at $A_{i} = 0.60$. The stability and morphology of the fingers are presented using the same color/shape of experimental data points as in figure \ref{p_Ca_A053}(a). Steady simulations are presented as solid black lines.}
\label{Pb_Ca_Ainf_060_appendix_C}
\end{figure}

Figures \ref{Pb_Ca_Ainf_082_appendix_C} and \ref{Pb_Ca_Ainf_060_appendix_C} show the results from experiments and steady simulations conducted at fixed levels of collapse $A_{i}=0.82$ and $A_{i}=0.60$, respectively. In each figure, part (a) shows the time evolution of the fingers for increasing values of $\overline{Ca}$ and part (b) the bubble pressure as a function of $\overline{Ca}$.
Finger propagation at $A_i =0.82$ is broadly similar to observations at $A_{i}=0.95$ described in \S \ref{slightly_collapsed_paper_2}, but here the fingers transition from round-tipped symmetric ($\overline{Ca}=0.51$) to flat-tipped asymmetric ($\overline{Ca}=0.90$). Furthermore, propagation is unsteady for the largest value of capillary number, $\overline{Ca}=1.08$. For $A_i=0.60$, this same transition between symmetric and asymmetric fingers is displaced towards lower values of mean capillary number $\overline{Ca}=0.16$. In the vicinity of this transition unsteady fingers can be observed with the development of deep clefts, but they become less pronounced as $\overline{Ca}$ increases. As we increase  $\overline{Ca}$  even further (panel 6 in figure~\ref{Pb_Ca_Ainf_060_appendix_C}(a)), the finger starts to develop deep indentations again, but with regular frequency. Finally, for even larger values of $\overline{Ca}$ we see the emergence of feathered modes. By contrast with $A_{i} = 0.53$, no steady pointed fingers have been observed in the experiments. The overall trend in the experimental data is that as the initial collapse is increased, smaller capillary numbers are required for the onset of complex unsteady behaviour.

\bibliography{references} \bibliographystyle{jfm}

\end{document}